\documentclass[fleqn,10pt]{wlscirep}
\usepackage[utf8]{inputenc}
\usepackage[T1]{fontenc}
\usepackage{subcaption}
\usepackage{datatool}
\usepackage[square, comma, numbers, sort&compress]{natbib}
\usepackage{multirow}
\usepackage{multicol}
\usepackage{longtable}
\usepackage{lscape}
\usepackage{lineno}
\newcommand{\Autoref}[1]{%
  \begingroup%
  \def\chapterautorefname{Chapter}%
  \def\sectionautorefname{Section}%
  \def\subsectionautorefname{Subsection}%
  \autoref{#1}%
  \endgroup%
}

\title{Synergistic Small Worlds that Drive Technological Sophistication}

\author[1,2,3,*]{Hardik Rajpal}
\author[3]{Omar A Guerrero}

\affil[1]{Centre for Complexity Sciences, Imperial College London, London, SW7 2AZ, United Kingdom}
\affil[2]{Department of Mathematics, Imperial College London, London, SW7 2AZ, United Kingdom}
\affil[3]{The Alan Turing Institute, London, NW1 2DB, United Kingdom}

\affil[*]{h.rajpal15@imperial.ac.uk}

\begin{abstract}
Advanced economies exhibit a high degree of sophistication in the creation of various products.
While critical to such sophistication, the nature and underlying structure of the interactions taking place inside production processes remain opaque when studying large systems such as industries or entire economies.
Using partial information decomposition, we quantify the nature of these interactions, allowing us to infer how much innovation stems form specific input interactions and how they are structured.
These estimates yield a novel picture of the nuanced interactions underpinning technological sophistication.
By analyzing networks of synergistic interactions, we find that more sophisticated industries tend to exhibit highly modular small-world topologies; with the tertiary sector as its central connective core.
Countries and industries that have a well-established connective core and specialized modules exhibit higher economic complexity and output efficiency.
Similar modular networks have been found to be responsible for maintaining a balance between integration and segregation of information in the human brain, suggesting a universal principle underlying the organization of sophisticated production processes.  
\end{abstract}
\begin{document}

\flushbottom
\maketitle
\thispagestyle{empty}
% \linenumbers

\section{Introduction}

Technological sophistication is central to the evolution of production processes.\footnote{By production process, we refer to the procedure through which a certain technology transforms a set of inputs into an output.}
For decades, researchers and scholars from different disciplines have worked towards deciphering the principles underpinning technological sophistication and to provide a general framework to reveal its structure, i.e., inferring how the different parts of a production process interact.
For instance, operations research tends to analyze supply chains and global value chains, human and economic geography focus on systems analysis, economics studies input-output models and fit production functions, systems engineering construct design structure matrices, and innovation scholars construct combinatorial models and perform network analysis on patent data. 
Despite these notable contributions, quantitatively inferring the nature of the interactions taking place in a production process and their structure remains largely unsolved.

This paper tackles these two challenges: (1) quantifying the nature of every interaction between the inputs of a production process, and (2) inferring the network structure of these interactions.
The former provides a novel measure of technological sophistication that explains economic complexity.
The latter reveals, for the first time, the nuanced interaction structure that takes place inside production processes and explains the economic sophistication of industries.

By leveraging partial information decomposition, we quantify the degree of synergy between the inputs of a production process.
In essence, the proposed method measures the contribution of an input interaction to the output; capturing not only input-input relations, but also inputs-output interdependencies.
To the extent of our knowledge, this is the first time that the nature of technology is inferred at such granular level and without assuming, \textit{ex ante}, analytic objects such as production functions.
We validate our synergy scores with different datasets and show that it predicts popular output-based indices that are often used as proxies of economic sophistication.
Then, we construct the synergy networks underlying different industries and analyze the topological features characterizing the structural principles of technological sophistication.

The synergy networks underpinning production processes suggest that more sophisticated industries exhibit small-world topologies.
However, they are a special class of modular small-world networks which use tertiary sector inputs acting as their connective core.
We find this fascinating as, in neuroscience, it has been shown that similar modular small-world structures support complex integration and consciousness in the brain, as described in the global neuronal workspace theory \citep{baars2007global, dehaene2001towards, shanahan2012brain, mashour2020conscious}.
Modularity appears to be crucial for technological sophistication alongside classic properties of small-world networks that enable global integration as well as functional segregation.
The ubiquity of these structures can be seen in a class of complex systems \citep{telesford2011ubiquity}, including social \citep{girvan2002community}, technological \citep{adamic1999small}, political \citep{tam2010legislative}, managerial \citep{uzzi2007small}, and biological \citep{jeong2001lethality} systems.

Our results suggest that sophisticated production processes face a similar challenge of balancing specialization in a diverse set of inputs as well as integrating them together to creatively generate new opportunities.
Thus, the proposed framework for quantifying the nature and structure of production processes is a major step towards opening the black box of technological sophistication and sheds new light on the micro-foundations of economic complexity.
In the rest of the paper we discuss the existing knowledge gap, describe the methods and datasets used, and present our results.

\section{Knowledge gap}\label{sec:literature}

In the theoretical literature of technological innovation, there is consensus agreeing on the principle that technological change can be described as a recombinant process through which novelty emerges from new ways of coupling existing devices (or technologies) to fulfill a certain purpose\citep{schumpeter_business_1939, henderson_architectural_1990, kauffman_technological_1995, weitzman_recombinant_1998, arthur_evolution_2006, arthur_structure_2007}.
This consensus is empirically supported by evidence on how new inventions recombine existing ones \citep{fleming_recombinant_2001, gittelman_does_2003, youn_invention_2015} (see \citep{savino_search_2017} for a comprehensive review).
Arguably, technological sophistication is intimately related to this transformative process, so producing empirical metrics to characterize it would provide generalizable foundations for its quantification.
Unfortunately, much of the empirical evidence in innovation studies remains domain-specific (e.g., patents \citep{sbardella_green_2018} and cities \citep{straccamore_urban_2022}), making it difficult to build generalizable empirical frameworks.
In our view, the lack of such frameworks obeys two limiting factors: (1) technology can be highly diverse across industries and countries, and (2) generating reliable estimates for production models with numerous inputs requires big data, something difficult to obtain from input-output (IO) tables (with the exception of recent developments of inter-firm transaction data \citep{diem_quantifying_2022}).

IO scholars and analysts circumvent the lack of big data by modeling production networks that assume, \textit{ex ante}, the nature of the input-input and input-output interactions through production functions \citep{leontief_structure_1941, diem_quantifying_2022}.
This approach has the benefit of shifting the problem of modeling technological sophistication from one of inferring structures to one of fitting parameters.
This shift comes with problems that have been recently discusses along the lines of misspecification \citep{zambelli_aggregate_2018}, estimation biases \citep{gechert_death_2019}, and aggregation artifacts \citep{iyetomi_paradigm_2012, constantino_modeling_2019}.
Thus, assuming (parametric) production functions instead of inferring interaction structures limits our capacity to quantify technological sophistication. 

The aforementioned limitations have become such an important issue that supply-chain surveys have been conducted in an attempt to determine the degree of dependence on certain inputs by certain industries (see the IHS Markit survey in \citep{pichler_forecasting_2022} and \citep{geodis_geodis_2017}).
Moreover, such information, has been incorporated in the new generation of IO models \citep{pichler_forecasting_2022, diem_quantifying_2022}.
Hence, a data-driven approach that would facilitate the inference of interaction structures in production processes would help alleviate some of these problems.

Systems engineering, has created an alternative approach to production functions by developing the concept of Design Structure Matrices (DSMs) \citep{steward_design_1981, browning_applying_2001}.
DSMs describe networks of interactions between the different components of a production processes.
They provide a comprehensive tool to represent technological sophistication, but they require substantial knowledge about the process itself.
Thus, DSMs build on a top-down approach that demands substantial \textit{ex ante} knowledge; making it difficult to scale up to more aggregate levels such as industries or sectors.
These aggregation levels are crucial for the design and implementation of country- or industry-wide innovation strategies and policy intervention.
A data-driven framework to infer DSMs, or some of their components, would complement this literature in important ways.

In recent years, the literature of economic complexity has emerged with new proposals on how to quantify technological sophistication \citep{hidalgo_product_2007, hidalgo_building_2009, tacchella_new_2012, tacchella_dynamical_2018} (see \citep{hidalgo_economic_2021} for a comprehensive review).
Building on existing literature on export diversification \citep{hirschman_paternity_1964, klinger_diversification_2006}, two dominant approaches have become the gold standard in this field: the Economic Fitness Index (EFI) \citep{tacchella_new_2012} and the Economic Complexity Index (ECI) \citep{tacchella_new_2012}.
While these frameworks are built on different theoretical foundations (see \citep{sciarra_reconciling_2020} for a rigorous comparison), both of them try to map export profiles into indices that capture different elements of economic sophistication.
In spite of this growing literature, the technological foundations of production sophistication remain mostly theoretical.

Evidently, quantifying the degree and structure of technological sophistication is a problem that pertains to multiple disciplines and has crucial implications in the design and implementation of economic and innovation strategies.
The lack of generalizable quantitative methods that address the production process explicitly poses a major barrier to advance research in these fields.
At the same time, it creates a void that maintains these various disciplines, to a great extent, disconnected.
Our contribution fills this gap and provides a general framework that facilitates the quantification of technological sophistication.
Such contribution can help to develop deeper insights on the fundamentals of technological sophistication.

\section{Framework and data}

Let us provide a succinct description of the methodology and the data employed in our analysis while leaving more specific details for \autoref{sec:methods}.
Broadly speaking, our method seeks to estimate the mutual information between pairs of inputs in a given industry (from IO tables).
Then, it decompose their contribution to the output into different information-sharing modes.
We focus on a particular mode known as \textit{synergistic information}, i.e., the information that cannot be obtained from any of the inputs alone as it exists as a virtue of their interactions.
Using this type of information, we produce a \textit{synergy score} capturing how complementary are certain inputs in a production process.

A higher synergy score means that input interactions produce more novel information.
Novelty, in turn, relates to the degree of sophistication of the output. 
This interpretation aligns with the established notion of a recombinant process generating novel outputs.
Thus, our synergy score provides a measure of technological sophistication that is explicit about the numerous interactions between inputs and outputs.

\subsection{The synergy score}

Consider an industry and three associated datasets.
Data $Y$ contain the changes in the output of the industry, while $X_1$ and $X_2$ capture the changes in inputs 1 and 2 respectively.
We are interested in quantifying how much information do $X_1$ and $X_2$ provide about $Y$.
This can be quantified using the total mutual information $I(X_1,X_2;Y)$ between the output and the inputs.
In essence, the total mutual information is a measure of the amount of uncertainty in $Y$, that is reduced by knowing $X_1$ and $X_2$.
Furthermore, it is possible to estimate how much of this uncertainty reduction is a result of the interactions between the inputs.
We can achieve this by following the partial information decomposition proposed by \citep{williams_nonnegative_2010}, where the total mutual information $I$ provided by  $X_1$ and $X_2$ about $Y$ can be decomposed as

\begin{equation}\label{eq:pid}
    %I(X_1,X_2;Y) = Syn(X_1,X_2;Y) + Oth(X_1,X_2;Y).
    I(X_1,X_2;Y) = Syn(X_1,X_2;Y) + Red(X_1,X_2;Y) + Unq(X_1;Y) + Unq(X_2;Y)
\end{equation}

In \autoref{eq:pid}, the mutual information is decomposed into the synergistic, redundant, and unique contributions (we explain the non-synergistic parts in \autoref{sec:methods}).
Synergistic information can only be obtained from the interaction between $X_1$ and $X_2$.
It is an analogue for input complementarity in the IO literature.
If either input is removed from the production process, all the synergistic information would be lost from the output signal.

More sophisticated production processes involve more synergistic information because they generate more novel outputs by recombining the same inputs in innovative ways.
Thus, we use this type of information as a synergy score and compute it for all unique pairs of inputs in the data.\footnote{\autoref{eq:pid} can be generalized for more than two inputs.
However, producing reliable estimates for more than three inputs can become data intensive.
Nevertheless, we show in the SI that our main results hold when we estimate synergies between triplets instead of pairs.}
Further details on the estimation method can be found in section \autoref{sec:info_theory}.

Using the pairwise synergy score for the inputs, we infer a weighted undirected network of the synergistic interactions in the production process. 
These synergy networks capture the structure of the technology underpinning a production process.
In contrast to the popular approach of assuming production functions in IO models, our synergy networks are inferred from the data.
Thus, they provide a `model-free' approach to quantify the degree and structure of technological sophistication.
Analyzing the topological properties of a large cross-section of these networks reveals features underpinning technological sophistication.

\subsection{Data}
\label{sec:workflow}

Our empirical application focuses on industrial technology through IO data, and places especial emphasis on the analysis of technologies with high heterogeneity in terms of their sophistication.
To empirically capture such diversity, it is necessary to assemble a dataset with substantial variation in terms of economic development.
Thus, having a large number of countries is key, as most technological variation would be expected between nations with different levels of development.
We find such coverage in the Eora26 dataset \citep{lenzen_mapping_2012}, a global collection of input-output tables with harmonized industries across a large number of countries.
The subset extracted from Eora26 contains annual input-output tables for 26 industries across 148 countries during the period 1995-2020.
We use the time series of each input and the corresponding total output to compute the synergy scores and networks associated with a particular country and industry.
Thus, we exploit the temporal variation in the data to build the scores, and then focus on comparing countries and industries.
The original time series are transformed into log-fluctuations (i.e., growth rates).\footnote{This transformation yields normally-distributed growth rates at the industrial level, which makes the data compatible with Gaussian estimators of mutual information.}
Further information on the data and its processing can be found in \autoref{sec:methods}.

\subsection{Validation}\label{sec:validation}

\Autoref{sec:literature} has discussed the emergence of economic sophistication indices in the economic complexity literature.
Three well-known measures in this community are the Economic Fitness Index (EFI) \citep{tacchella_new_2012}, the Economic Complexity Index (ECI) \citep{hidalgo_product_2007}, and a generalization of the previous two, the GENEPY index\citep{sciarra_reconciling_2020}.
While commonly used on international trade data, these indices can also be calculated at the industry level.\footnote{To the extent of our knowledge, across the now sizable literature on economic complexity, all studies focus on national and regional analysis, but there are no industry-level applications.
This task requires mapping products into industries, which we do and explain in the SI.}
As these indices are well-accepted proxies of how sophisticated is an economy, we validate our method by showing how our synergy score is able to predict these three measures.
Note that, to compute these indices, we employ data that are independent from the aforementioned IO tables: the BACI international trade dataset.

\begin{figure}[ht]
\centering
\caption{Synergy score predicting export-based industry sophistication indices}\label{fig:main_result}
    \begin{minipage}{0.32\textwidth}
        \subcaption{Economic fitness index}\label{fig:main_result.fitness}
        \includegraphics[angle=0,width=1.\textwidth]{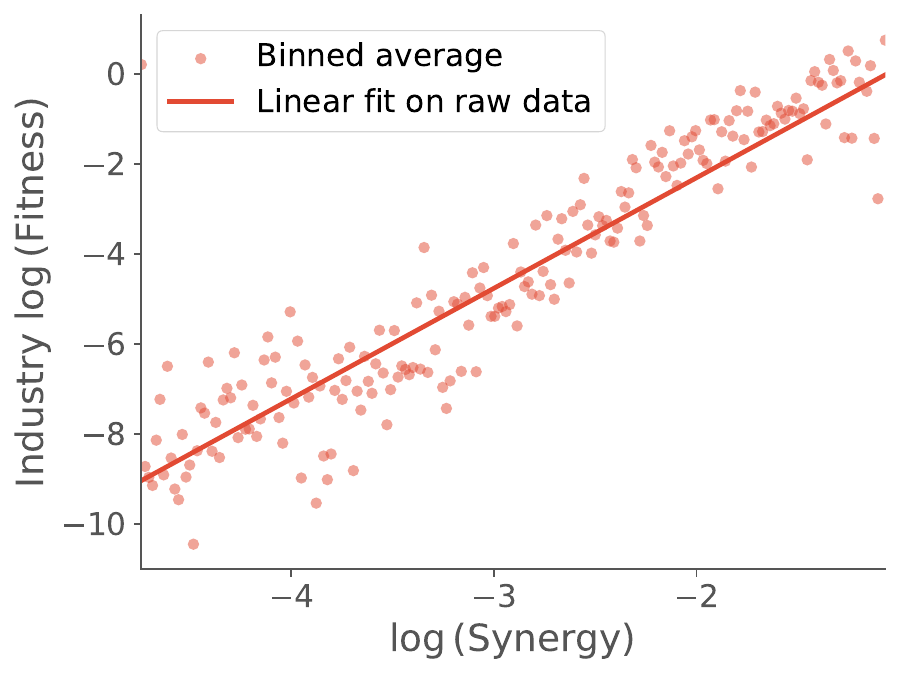}
    \end{minipage}
    \begin{minipage}{0.32\textwidth}
        \subcaption{GENEPY index}\label{fig:main_result.genepy}
        \includegraphics[angle=0,width=1.\textwidth]{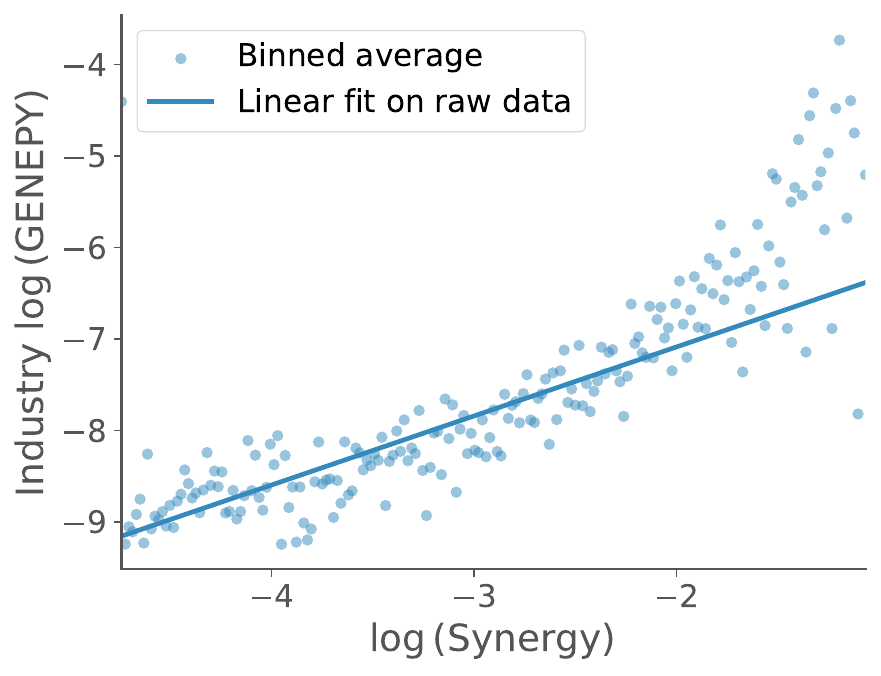}
    \end{minipage}
    \begin{minipage}{0.32\textwidth}
        \subcaption{Economic complexity index}\label{fig:main_result.complexity}
        \includegraphics[angle=0,width=1.\textwidth]{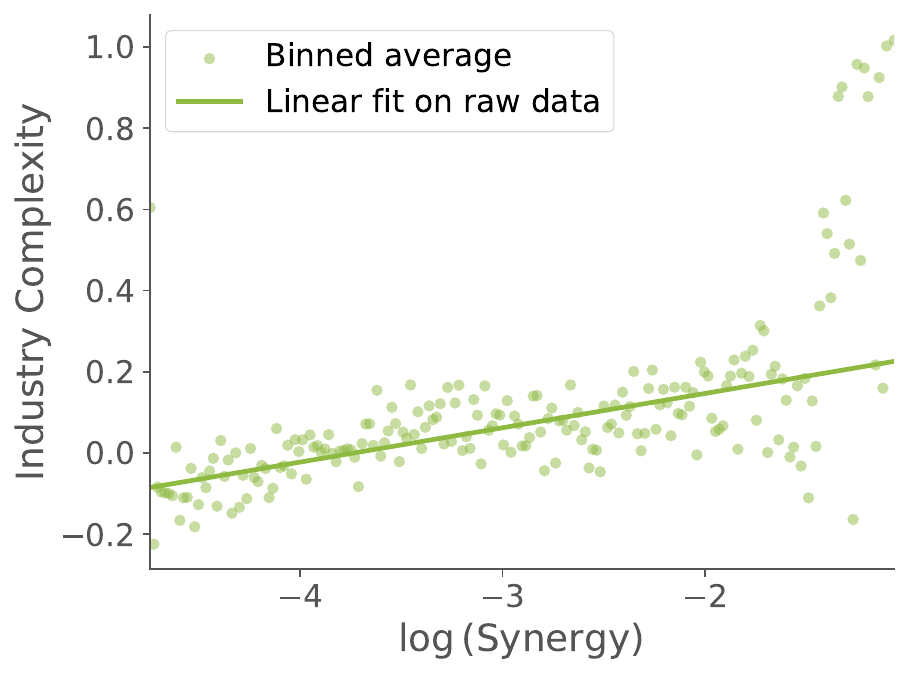}
    \end{minipage}
\caption*{\footnotesize 
\textit{\textbf{Notes}}: With the purpose of a clear visualization, we display data binned according to the synergy scores.
Each dot corresponds to the average value withing the corresponding bin.
Since the synergy distribution is right-skewed, we analyze the logarithm of the scores.
The linear fit is estimated using the entire data (not the binned one).}
\end{figure}

\autoref{fig:main_result} shows a positive correlation between the synergy score and the three indices.
It suggests that synergistic interactions predict these proxies of economic sophistication, validating our score.
Let us reiterate on the fact that, while we infer technological sophistication by directly analyzing input-input and input-output interactions, complexity indices take an approach that focuses on the co-occurrence of outputs (so input-output interactions are not taken into account).
We find it remarkable that our method is able to capture the cross-country and cross-industry variation produced by these indices, especially when these two distinct approaches use independent datasets.

\hspace{5cm}

In the SI, we provide more formal validation tests using different regression models that control for factors such as GDP per capita and energy efficiency.
Furthermore, we extend these validation tests to different levels of aggregation, alternative ways of computing the synergy scores, and alternative IO data.
Together, these analyses provide strong support to our framework.

\section{Results}
\label{sec:results}

One of the main advantages of our approach is the explicit calculation of synergy scores for every interaction between inputs.
This means that, when analyzing pairwise interactions, it is possible to construct an undirected network with weights determined by the synergy scores.
These synergy networks capture the structure of the technology underpinning a production process.
The details on how to construct synergy networks can be found in \autoref{sec:info_theory}.

\begin{figure}[ht]
    \centering
    \caption{Synergy networks of the transport industry}\label{fig:community}
    \includegraphics[width=0.7\textwidth]{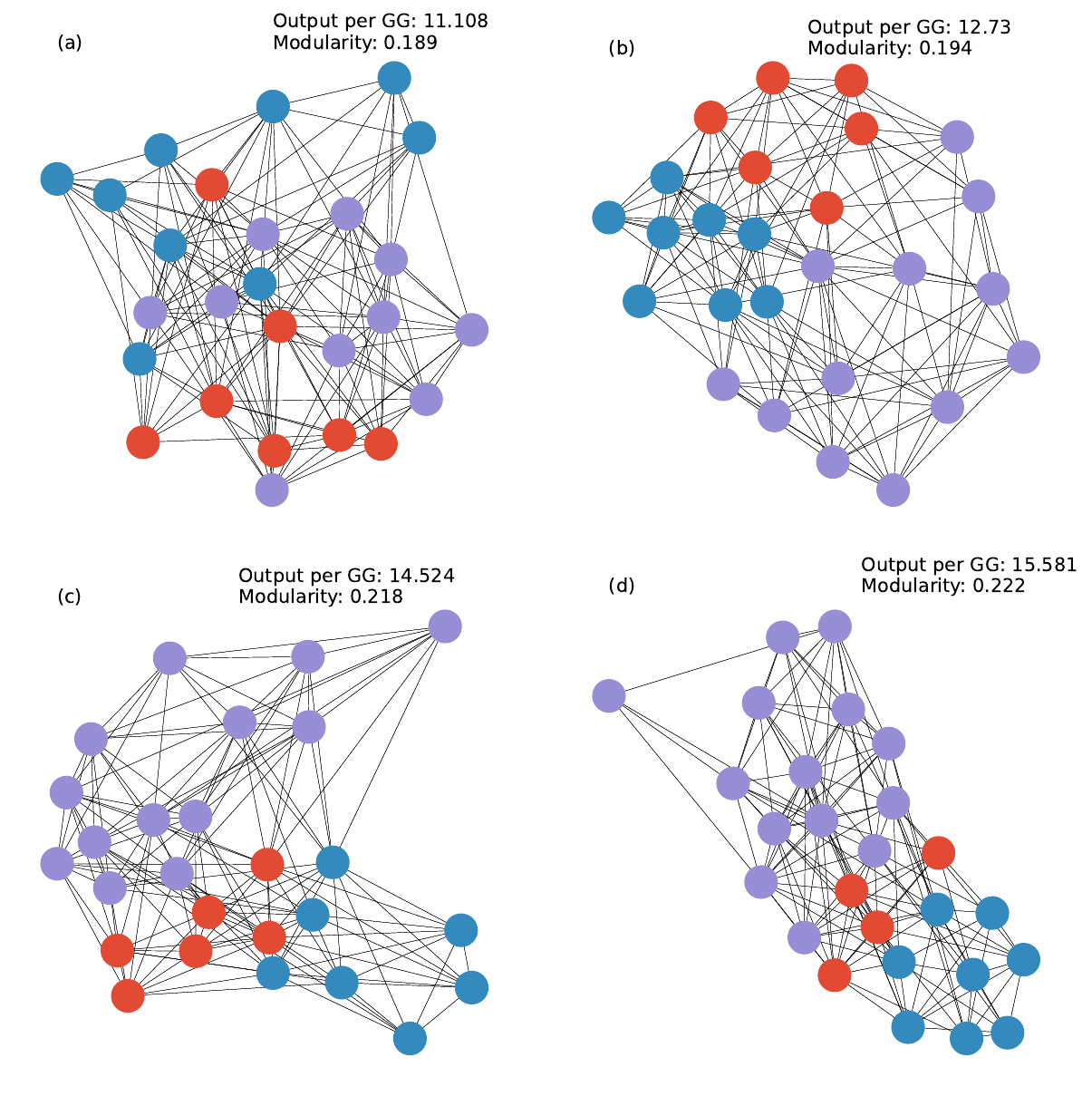}
    \caption*{\footnotesize 
\textit{\textbf{Notes}}: The node colors represent community labels.
GG stands for greenhouse gas.}   
\end{figure}

First, let us provide an example of one such network and how its structure differs across countries with different levels of technological sophistication.
\Autoref{fig:community} presents the case of the transport industry.
Arguably, technologies with a lower rate of greenhouse gas (GG) emissions per unit output are more sophisticated.
The plot displays four country clusters, ordered from the one with the least sophisticated technology (denoted by a larger output per GG emission) to the most advanced one.
The nodes in each network correspond to industries (including transport itself) that sell inputs to the transport industry.

To better illustrate the structure of these networks, we apply the Louvain algorithm to detect three communities and display them using spring layouts.
The nodes are colored according to each community.
The first thing to notice is that, the higher the output per GG emission, the more separated are the communities.
This is confirmed by higher modularity scores.
Second, the separation between communities is not random, as the red nodes seem to play a brokerage or intermediary role when technology is more sophisticated.
This is fascinating as it would point towards a specialization vs generalization trade-off in technological structure as an industry becomes more sophisticated.
Furthermore, the potential existence of intermediary industries begs the question of which ones tend to be those?
We investigate these questions in a more systematic way and present our findings in the remainder of this section.

We analyze the topological properties of 100+ synergy networks inferred through our method.
First, let us investigate the question of which are the industries that tend to be intermediaries.
As the specific nodes with a brokerage role may vary between clusters and industries, we resort to a broader categorization based on the economic sector to which each input belongs to: primary, secondary, and tertiary.
Then, we estimate the mixing probabilities between sectors and perform a T tests for these association probabilities.\footnote{With reference to a null model (described in the SI), followed by a false-discovery-rate correction for multiple comparisons.}

\begin{figure}[ht]
\centering
\caption{Sector differences in network properties}\label{fig:sector_diff}
    \begin{minipage}{0.49\textwidth}
        \subcaption{Sector mixing probabilities}\label{fig:net_mixing}
        \includegraphics[width=\textwidth]{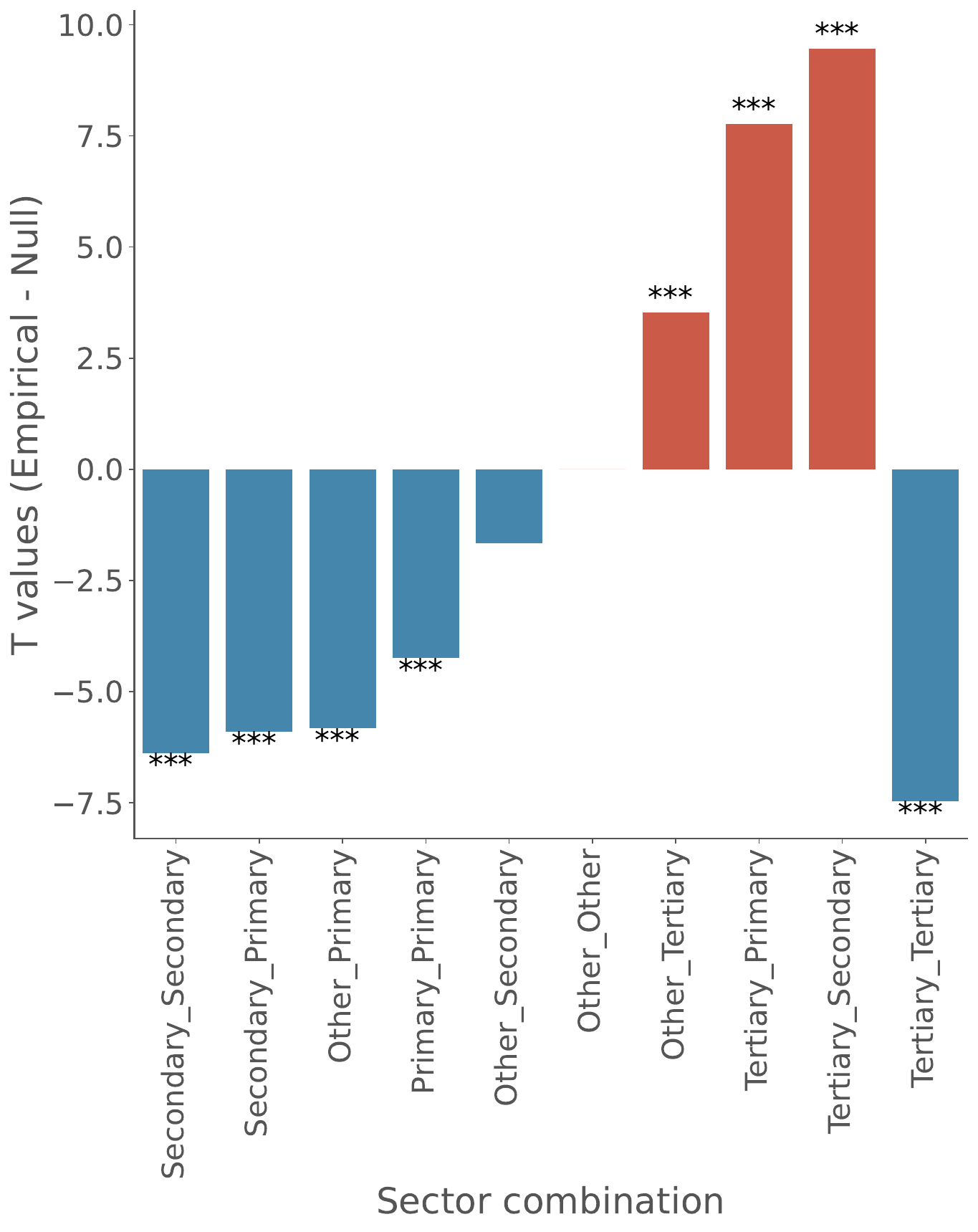}
    \end{minipage}
    \begin{minipage}{0.49\textwidth}
        \subcaption{Sector measures of hub-ness}\label{fig:sector_hub}
        \includegraphics[width=\textwidth]{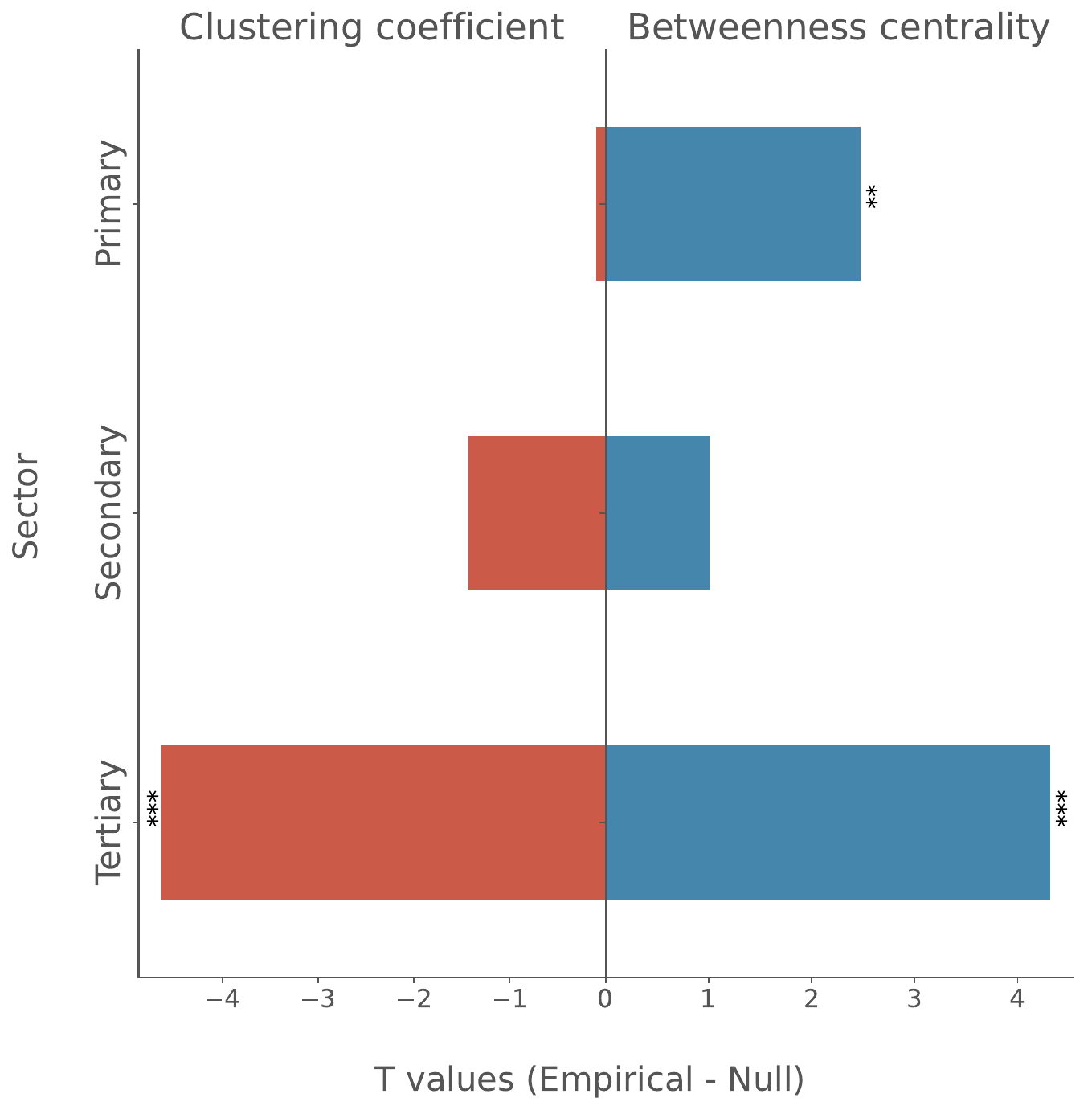}
    \end{minipage}
    
    \caption*{\footnotesize{
    \textit{\textbf{Notes}}: These estimates are produced by computing the difference in the relevant metric between the empirical networks and the ones produced by the null model.
    Confidence levels are indicated by * for 90\%, ** for 95\%, and *** for 99\%.}}
\end{figure}

\Autoref{fig:net_mixing} presents the T values of every pair of sectors.
As we can see, the tertiary sector interacts substantially more with the other two sectors.
This suggests that industries in the tertiary sector (such as communication, hospitality, finance, education) tend to mediate synergies between industries from the other sectors of the economy. 

For a robustness test, we estimate two `hub-ness' measures and corroborate if our findings regarding the tertiary sector are consistent.
Hubs are usually characterized by high betweenness centrality and low clustering in networks.
\Autoref{fig:sector_hub} suggests that this is indeed the case for industries in the tertiary sector, distinct from those in the primary and secondary sectors (as compared to a null model).

\hspace{5cm}

The previous analyses confirm that a subset of inputs tend to mediate the production process of industries.
Furthermore, such industries are classified into the tertiary sector.
These findings are consistent with the accepted idea that membership to a particular sector conveys information about structural differences between industries and their underlying technologies, as they correspond to the main stages of production \citep{fisher_production_1939, clark_revised_1940}.
Thus, our findings not only reveal structural features about technological sophistication, but also confirm long-argued ideas about industrial development.

\begin{figure}
    \centering
    \caption{Conditional correlations of salient factors with output efficiency}
    \begin{minipage}{0.66\textwidth}
        \subcaption{Factor loadings}\label{fig:net_factor}
        \includegraphics[width=\textwidth]{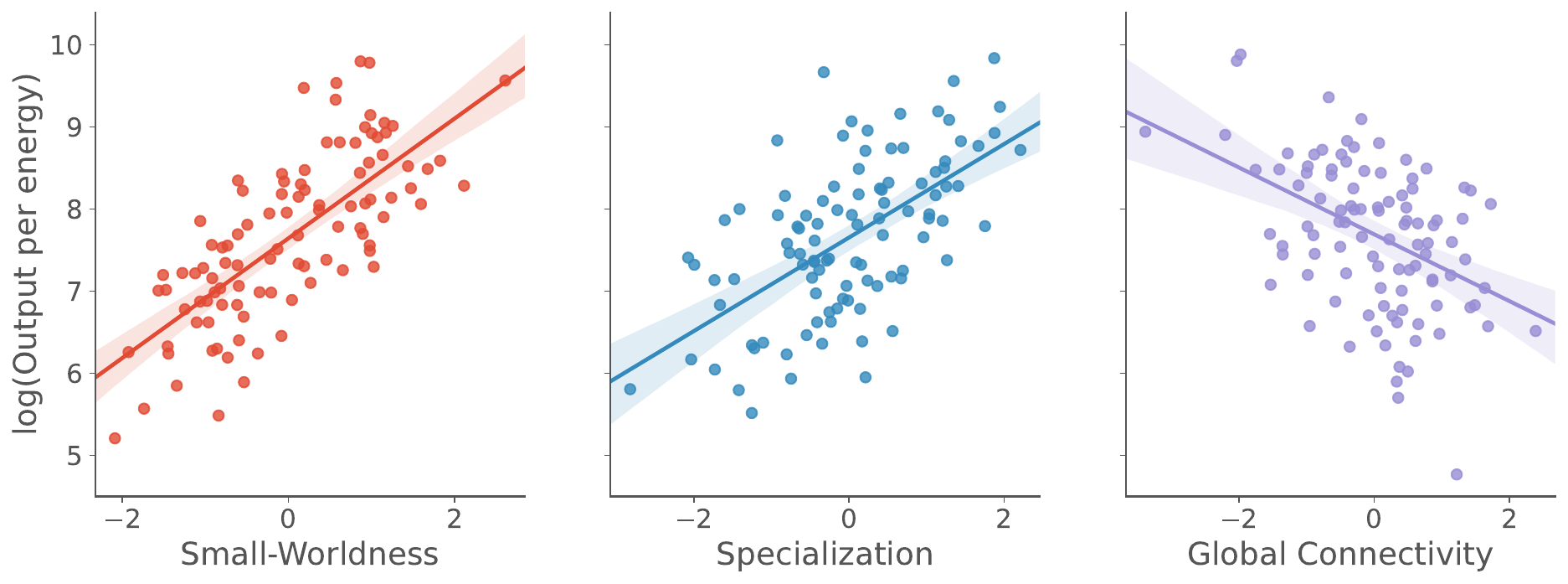}
    \end{minipage}
    \begin{minipage}{0.33\textwidth}
        \subcaption{Factor-output correlation}\label{fig:corr_factor}
        \includegraphics[width=\textwidth]{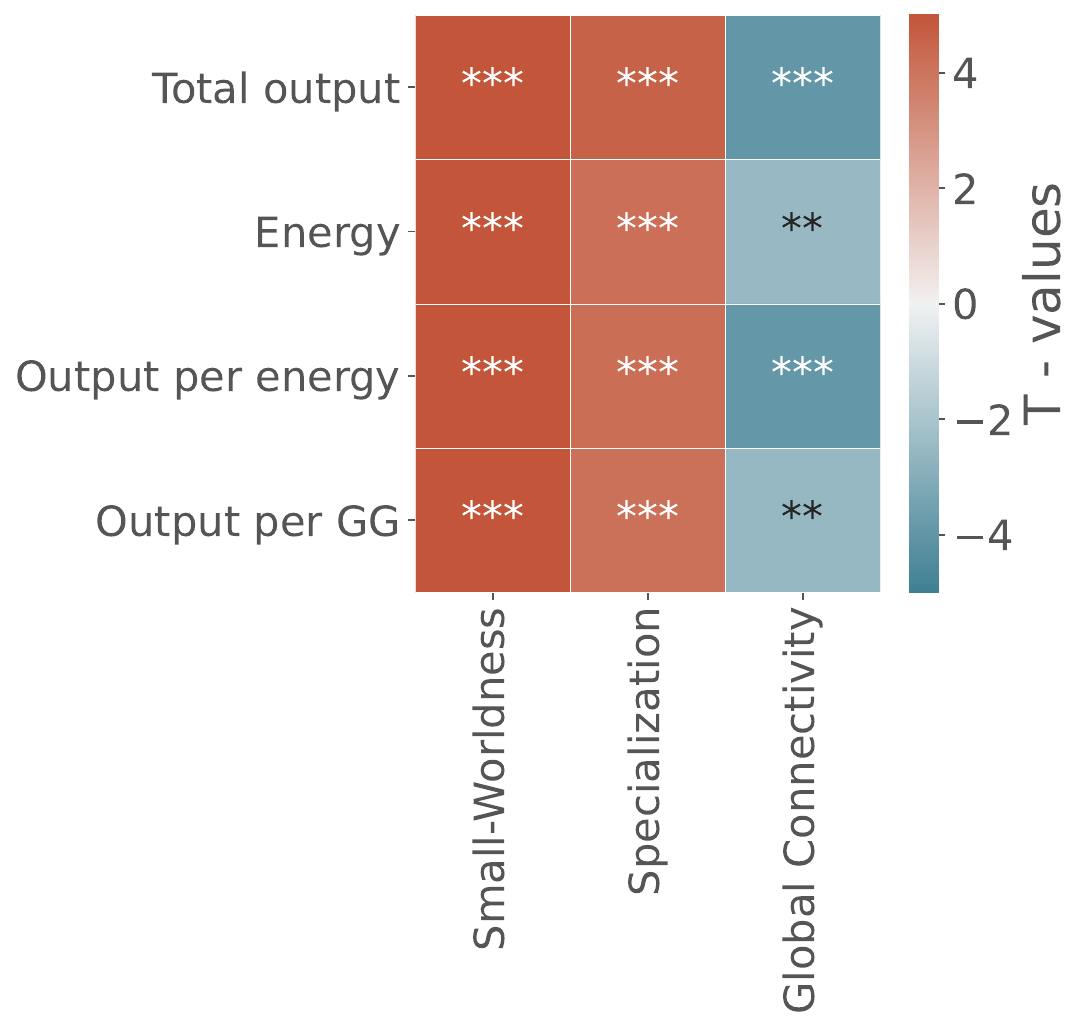}
    \end{minipage}
    \label{fig:net_factor_out}
        \caption*{\footnotesize{
    \textit{\textbf{Notes}}: Output per GG stands for greenhouse gas.}}
\end{figure}

Now, let us analyze a broader set of (global) topological properties in synergy networks to uncover those features that underpin technological sophistication.
To this end, we estimate 14 network measures that can be classified into three categories of network metrics: small-worldness, specialization, and global connectivity.
Since several of these measures tend to correlate under certain topologies \citep{li2011correlation}, we perform factor analysis to disentangle their contributions to each one of the three categories of network metrics.\footnote{The full list of measures and their contribution to specific factors is provided in the SI.}
We find that these factors significantly correlate with various output metrics that denote more sophisticated production processes, namely, total output, energy consumed, output per unit energy consumed, and output per unit GG emitted, as suggested by \autoref{fig:net_factor_out}.

Using a robust linear model, we predict the four output measures with the leading factors.
As observed in \autoref{fig:net_factor}, small-worldness and specialization are positively correlated with the output metrics.
In contrast, global connectivity exhibits a negative correlation.
Since, global connectivity relies on measures like algebraic connectivity, it can be inferred that the productive structures enabling more sophisticated processes are more fragile; they also rely on a few key nodes to keep the sub networks connected.

\autoref{fig:corr_factor} suggests that a higher output per greenhouse gas emissions correlates with more modularity.
In contrast, low output per emissions associates with a less modular structure overall.
In the SI, we show that this characteristic structure is not directly observable in IO data without information-decomposition analysis in this paper.
Neither are the relationships between network structure and output efficiency documented here.

\section{Discussion and conclusions}

The quantification of technological sophistication in production processes is an elusive problem, relevant to several disciplines.
This paper introduces the first method that explicitly addresses input-input and input-output interactions, opening the black box of production processes.
By estimating the amount of synergistic mutual information between the inputs of a production process, we quantify the degree of technological sophistication across various industries and countries.
Moreover, we infer the structure of these technologies by constructing synergistic interaction networks, revealing features that characterize industrial sophistication.
These networks provide empirical grounds to select and justify production functions in IO models; something missing in this large body of literature and a major limitation in IO empirical studies.
They also reveal the structural role of industries with a high degree of synergistic interactions.
Finally, they suggest certain universality in the prevalence of modular small-world topologies across various classes of complex systems that perform sophisticated behaviors.

Some of the main limitations of this approach are that it requires larger data than what is typically found in IO tables.
This means that, in this study, we need to develop a clustering procedure (see the \autoref{sec:methods}), and that our inferences are not for specific countries, but for groups.
Another limitation is that we have to sacrifice the temporal dimension as we need to exploit time variation to produce the estimations.
Ideally one would have high-frequency IO tables to compute the synergy scores by sub-periods.
This would allow us to study, for example, the evolution of technological sophistication and of its networked structure across countries and industries.
Fortunately, new firm-transaction datasets are being generated as we write this manuscript, so the future for using the proposed approach looks very promising.
Furthermore, because our framework works on the premise of a generic production process its applications could extend to other domains such as the study physical/biological systems.

\section{Methods}\label{sec:methods}

\begin{figure}
    \centering
    \caption{The two-step workflow}
    \label{fig:workflow}
    \includegraphics[angle=0,width=1\textwidth]{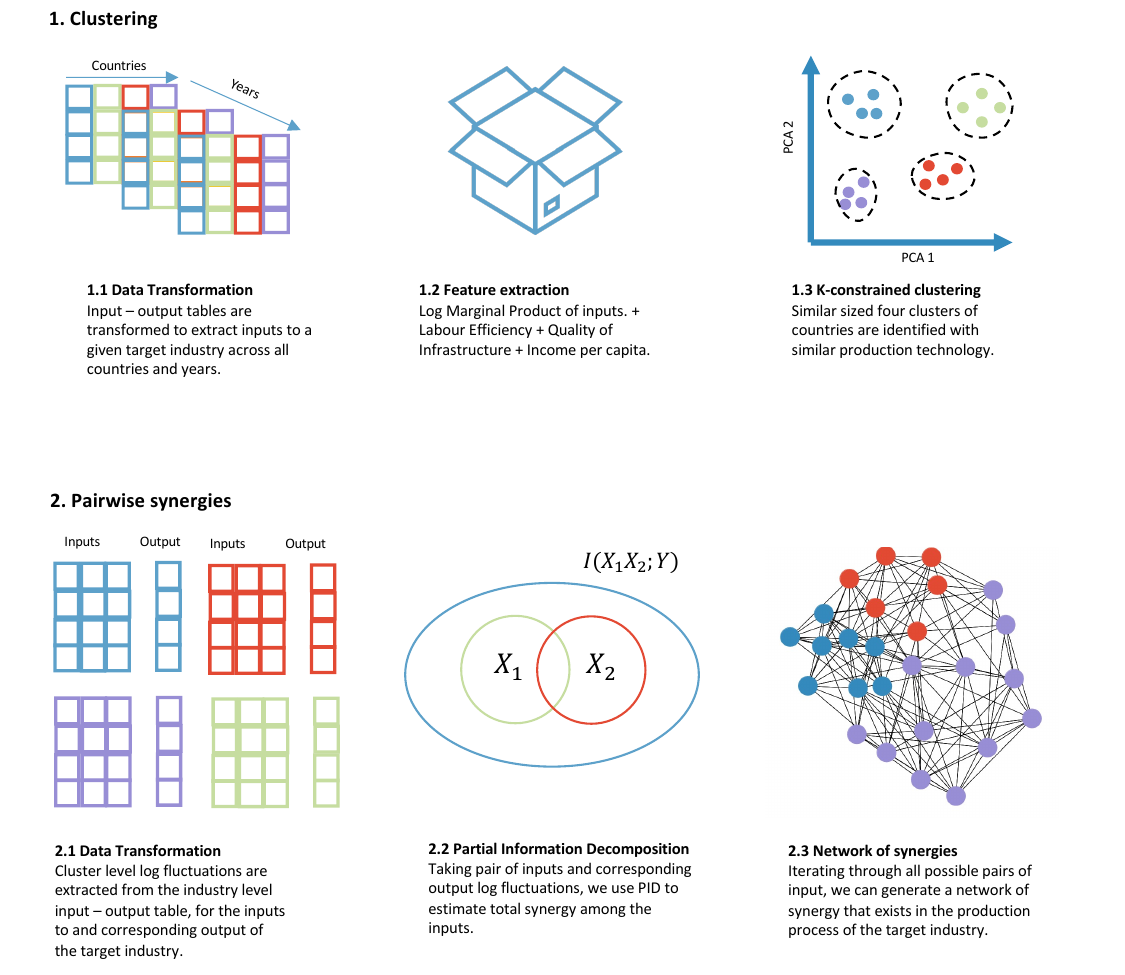}
% \caption*{\footnotesize 
% \textit{\textbf{Notes}}: ...}
\end{figure}

Partial information decomposition (PID) requires estimating the joint probability distributions of inputs and outputs.
To do this robustly and efficiently, a sufficiently large number of observations is required (the number of required observations scales with the number of inputs).
Such data do not exist for a single industry in a country's IO tables.
Nevertheless, we can overcome this limitation by performing a data-clustering procedure based on the technological similarity of countries in a given industry.
Thus, we create a workflow that facilitates the pre-processing and inference tasks.
The entire pipeline should be repeated for each industry that one would like to include in the analysis.
The workflow consists of two major steps, and an illustrative sketch is provided in \autoref{fig:workflow}.

\begin{enumerate}
    \item \textbf{Clustering}: Grouping countries that exhibit similar production patterns in a target industry, preferably in roughly equally sized groups.
    This means that two countries that are in the same cluster in a given industry A, may be in different ones when analyzing another industry B.
    \item \textbf{Estimation}: Using PID to estimate synergy scores for each pair of inputs in the target industry, and constructing its corresponding synergy interaction network.
\end{enumerate}

\hspace{5cm}

In the rest of this section, we explain the PID method and the clustering procedure in detail.
Since the economic complexity indices are standard metrics in the literature, we explain how we calculate them at the industry level in the SI.

\subsection{Synergy score}
\label{sec:info_theory}

% Since the seminal paper of Claude Shannon in 1948 \citep{shannon_mathematical_1948}, information theory has evolved substantially, spanning beyond the study of communication.
% It has become a powerful tool to quantify interdependencies between variables in many complex systems like the ecology \citep{ulanowicz_information_2001}, financial markets \citep{zhou_applications_2013}, the brain \citep{dimitrov_information_2011} and thermodynamics \citep{parrondo_thermodynamics_2015}.
Shannon entropy measures the variability of a system in a given state-space.
Leveraging this concept, Shannon proposed a measure of dependence between two variables, commonly known as mutual information.
The ability to assess the variability of a system--and the interdependency among its variables--made information theory ideal for the empirical study of complex systems.   
We adapt these concepts and tools to quantify the interdependencies between an industry's inputs through a synergy score.

\subsubsection{Mutual information}

% Consider a random variable $X$, defined on a discrete state space $\mathcal{X}$.
% The Shannon entropy is written as

% \begin{equation}
%     H(x) = -\sum_{x \in \mathcal{X}} p(x) \log p(x),
% \end{equation}
% where $p(x)$ is the probability of the realization $x$ of variable $X$ on the state space $\mathcal{X}$.
% Entropy is highest when the variable explores all possible states with equal probability.
% Similarly, 

The mutual information between two random variables $X$ and $Y$ can be defined as below,

\begin{equation}
    I(X;Y) = \sum_{y \in \mathcal{Y}} \sum_{x \in \mathcal{X}} p(x,y) \log\left(\frac{p(x,y)}{p(x)p(y)}\right),
\end{equation}
where $p(x)$ is the probability distribution corresponding to the random variable $X$ on the state space $\mathcal{X}$ and $p(x,y)$ represents the joint distribution of the two random variables.
The measure $I(X;Y)$ becomes 0 if $X$ and $Y$ are independent, such that $p(x,y) = p(x)p(y)$.

For the case of gaussian random variables, the corresponding mutual information can be written as (see\citep{mackay_information_2003} for a complete derivation),
\begin{equation}
    I(X;Y) = \frac{1}{2}\log\left(\frac{det \Sigma(X)}{det\Sigma(X|Y)}\right),\label{eq:gaussian_mutual}
\end{equation}
where $det\Sigma(X)$ represents the determinant of the covariance matrix of $X$, and $\Sigma(X|Y)$ represents the conditional covariance, which can be written as

\begin{equation}
    \Sigma(X|Y) = \Sigma(X) - \Sigma(X,Y)\Sigma(Y)^{-1}\Sigma(Y,X).
\end{equation}

The definition in \autoref{eq:gaussian_mutual} also exists for a multi-variate setting.
For three or more variables, higher-order effects such as synergistic interactions can be observed as well using the PID framework\citep{williams_nonnegative_2010}. 
% Interestingly, these synergistic interactions are present in the groups of variables (inputs) as a whole, so they are absent when considered individually. 
%  formalize the measurement of synergy for the case with two inputs and one output using PID. 
% Thus, we proceed to briefly explain the PID framework.

\subsubsection{Partial information decomposition}

PID decomposes the mutual information between a pair of inputs and the output into \textit{Synergistic}, \textit{Redundant} and \textit{Unique} information.
This decomposition was first introduced in \citep{williams_nonnegative_2010}, and has been instrumental to study different types of interactions between random variables \citep{lizier_information_2018,wibral_partial_2017,barrett_exploration_2015}. 

\begin{figure}
    \centering
    \caption{Partial information decomposition}
    \label{fig:pid_venn}
    \includegraphics[width=0.75\textwidth]{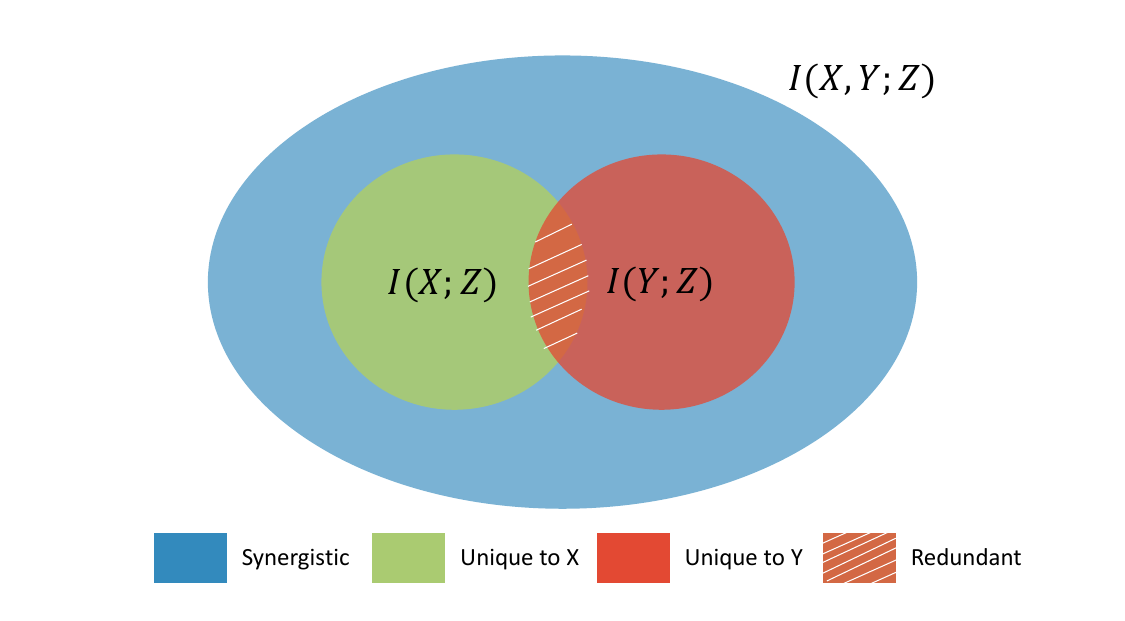}
\end{figure}

Let us look at the case of mutual information between two input variables ($X_1, X_2$) and one output variable ($Y$).
Here, the total mutual information about the output provided by the two inputs can be represented as the following Venn diagram\autoref{fig:pid_venn}.

\hspace{5cm}

Redundant information could come from either $X_1$ or $X_2$, so this information would be preserved if one of the inputs was removed.
It is an analogue of the concept of input substitutability from the IO literature.
Unique information comes from each input only, so it represents the unique contribution that each input makes to the output.
Both of these types of information are contained in the term $Oth(X_1,X_2;Y)$ from \autoref{eq:pid}.
Thus, let us rewrite it in its extended form

\begin{equation}\label{eq:pid_extended}
    I(X_1,X_2;Y) = Syn(X_1,X_2;Y) + Red(X_1,X_2;Y) + Unq(X_1;Y) + Unq(X_2;Y).
\end{equation}

A general formulation of \autoref{eq:pid_extended} for $n$ random variables is provided by \citep{williams_nonnegative_2010}.\footnote{Performing the PID for more joint variables demands substantially more observations.
To demonstrate the robustness of our pairwise results, the SI shows that our findings hold when using triplets instead of pairs.}
The unique and the total joint information about the output provided by the inputs can be exactly calculated.
However, to estimate the synergy and redundancy between the inputs, we need to assume a redundancy function. 
Here, we employ the minimum mutual information redundancy estimator developed for multivariate Gaussian systems \citep{barrett_exploration_2015}.
This redundancy estimator assumes the total redundancy to be equivalent to the unique mutual information about the output provided by the weakest input.
Formally, this is

\begin{equation*}
    Red(X_1,X_2;Y) = \min(I(X_1;Y),I(X_2;Y)).
\end{equation*}

Following \citep{barrett_exploration_2015}, using this redundancy function yields the estimator of the synergistic interaction to be

\begin{equation*}
    Syn(X_1,X_2;Y) = I(X_1,X_2;Y) - \max(I(X_1;Y),I(X_2;Y))
\end{equation*}

Note that $Syn(X_1,X_2;Y)$ is a distance measure because mutual information can be written in terms of the Kullback-Leibler (KL) distance between two random variables.
This means that the synergy score is comparable across the industries and countries in our study because all the production processes have the same number of random variables, providing the same bounds to their KL distances.
In terms of units, $Syn(X_1,X_2;Y)$ can be encoded in bits to obtain a more concrete measurement of how much synergistic information represents from the total mutual information.

To remove potential numerical artifacts, we implement a bias correction procedure by performing a randomized estimation where the inputs are shuffled.
The synergy score calculated using the randomized data is then subtracted from the true synergy score.
This ensures that the statistical significance of the estimated mutual information by removing the effect of spurious correlations.

\subsubsection{Synergistic interaction networks}

The synergistic interaction network of a given industry is constructed using the synergy scores of every pair of inputs.
Each node in this network represents an industry (including itself) providing an input.
The edges have no direction and are weighted according to the pairwise synergy score.
Each network has 26 nodes.

It is possible to obtain fully connected graphs, which may trivialize certain network analysis.
Thus, we extract the backbone of each network by using the method discussed in \citep{coscia_network_2017}, which builds on the well-known disparity filter \citep{serrano_extracting_2009} by automating the selection of a filtering threshold.
To estimate the measures of centrality, clustering, and mixing probabilities discussed in figure \autoref{fig:net_mixing}, we use the binary version of these backbone networks.

When performing statistical analysis in \autoref{fig:sector_diff} on the synergy networks, we use a null model.
This model generates null networks by shuffling the edges across all possible combinations of nodes.
In this way, the null networks have a similar edge-weight distribution as the reference networks.
The shuffled structure allows for all possible inter-sector interactions.
One thousand null networks are generated for every reference network to correct for sampling bias.
The properties of interest such as the mixing probability, clustering, and betweenness are calculated both on the reference and the null networks.
The differences between the properties in these two distributions is used in the T tests.

\subsection{Data and prepossessing}
\label{sec:dataset}

\subsubsection{Main datasets}\label{sec:data}

\paragraph{Input-output tables:}
The IO data are obtained from the Eora26, constructed by \citep{lenzen_mapping_2012, lenzen_building_2013}.
These data come from IO tables from organizations such as the UN, Eurostat, and the OECD, among others.
It is the largest input-output dataset in terms of country coverage (181 economies).
It consists of 26 harmonized industries.
Eora26 has become standard in the study of global value chains \citep{unctad_transnational_2019} and material footprint \citep{wiedmann_material_2015}.
It has also been shown to be consistent with similar, but smaller, global databases \citep{moran_convergence_2014}.

\paragraph{International trade:}
In our validation, we calculate the indices of economic sophistication using the BACI international trade dataset \citep{gaulier_baci_2010}, which is independent from Eora26.
These data contain trade records between 200 countries for more than 5000 products during the 1995-2020 period.
These products are uniquely classified into 15 of the 26 industries of Eora26 through the HS92-to-ISIC correspondence tables of the UN Statistics Division.
% While there are alternative trade datasets BACI has been the only one so far used to measure economic complexity at the level of industries \citep{pugliese_economic_2021}, we use it as our primary source for estimating industry sophistication.

\paragraph{Development indicators:}
We use three auxiliary development indicators to improve our clustering procedure (see \autoref{sec:clustering}).
The first is the World Bank's gross national income (GNI) per capita indicator.
The second and third are the labor efficiency and infrastructure indicators from the World Economic Forum's Global Competitive Index Report.
The GNI covers the same period as Eora26, while the other two indicators are available from 2007-2017.\footnote{We find that this auxiliary information is very useful to obtain coherent technology clusters without trivially becoming the leading feature.}

\subsubsection{Preprocessing input-output data}

Using the Eora26 input-output tables, we construct time series for the total output of each industry (in a specific country), as well as for the total inflow (combining both domestic and foreign) of each of its inputs.
Formally, let $T_{i,j}^{c',c}$ denote the total transaction value from industry $i$ in country $c'$ to industry $j$ in country $c$ (in USD basic prices).
Then, the total input inflow from industry $i$ to $j$ for country $c$ is, 
\begin{equation}
    X_{i,j}^c = \sum_{c'} T_{i,j}^{c',c}.
\end{equation}

Similarly, the total output ($Y_j$) of industry $j$ in country $c$ can be defined as

\begin{equation}
    Y_j^c = \sum_{c',i} T_{j,i}^{c,c'} + \sum_{c'} F_{j}^{c,c'},
\end{equation}
where $F_{j}^{c,c'}$ represents the final demand of goods and services of sector $j$ of any country $c'$ in country $c$. 

By using the multi-region transaction matrices $T$ and $F$ we identify the total inter-industry inputs $X_{i,j}^c$ and the total output $Y_j^c$ of industry $j$ in country $c$.
These time series may be affected by temporal trends that do not exhibit a stationary distribution.
Thus, we convert these flows into log-fluctuations, a common practice in the study of financial time series.
The log-fluctuations for the input $i$ to industry $j$ and country $c$ ($X_{i,j}^c(t)$), and the corresponding output ($Y_j^{c}(t)$) can be written as

\begin{equation}
\begin{split}
    \hat{X}_{i,j}^{c}(t) &= \log \frac{X_{i,j}^{c}(t)}{X_{i,j}^{c}(t-1)} \\
    \hat{Y}_{j}^{c}(t) &= \log \frac{Y_j^c(t)}{Y_j^{c}(t-1)}.
\end{split}
\end{equation}

For a given industry, a vector of output (or input) log-fluctuations can be collated across years and countries for data-augmentation purposes.
This is necessary in the application of PID since the country-specific vectors $\hat{X}_{i,j}^{c}(t)$ and $\hat{Y}_{j}^{c}(t)$ are not long enough to fulfill the requirements of the data estimation method \citep{barrett_exploration_2015}.
Ideally, such collation should consider clustering countries with similar technologies in a given industry.
Thus, in the next section, we explain how to achieve this.

\subsection{Clustering similar technologies}\label{sec:clustering}

For a target industry, we cluster countries with similar technologies to augment the size of the data.
We measure technological similarity by using a popular concept in economics: the marginal product.
The marginal product is the relative change in the output of an industry with respect to change in one of its inputs. 
This measure enables us to dissect the effect of each input at a first level of approximation.
The marginal product of the input coming from industry $j$ in industry $i$ (in country $c$) is

\begin{equation}
    MP_{i,j}^c(j) = \frac{Y_j^{c}(t) - Y_j^{c}(t-1)}{X_{i,j}^{c}(t) - X_{i,j}^{c}(t-1)}.
\end{equation}

The median marginal product for all the available years, is taken to be a feature of the input to a particular industry.
This quantity usually exhibits a fat-tailed distribution across countries because of the characteristic output differences among the economies.
Thus, we log-transform it.

We use log $MP$ as the key set of features for clustering countries with similar production technologies, along with the auxiliary development indicators described in \autoref{sec:data}.
The indicators on GNI, labor efficiency, and infrastructure help avoiding trivial clusters that could result from the coincidental similarity of marginal products due to non-technological reasons such as exogenous shocks.

Finally, we employ the k-means-constrained clustering algorithm \citep{bradley_constrained_2000} to define four country clusters in a target industry.\footnote{For the OECD data, the optimal number of clusters is 3.}
We choose this constrained version of the k-means algorithm because it allows finding clusters that are balanced in size.\footnote{In the SI, we show that our results are robust to the traditional unconstrained k-means algorithm.
However, we prefer the constrained version as unbalanced clusters may introduce biases in the synergy scores due to under or over representation of certain types of countries.}
The hyperparameters of the algorithm, including the number of clusters, are optimized using consensus clustering.\footnote{The SI explains how principal component analysis of the feature matrix is used to ensure that the marginal product is the leading feature.}
\nolinenumbers
\bibliography{references}

\section{Acknowledgements}

We like to thank the AI for Science and Government (ASG) and Engineering and Physical Sciences Research Council (EPSRC) UK, for supporting this research as part of the Shocks and Resilience project at the Alan Turing Institute.

\section{Author contributions statement}

HR and OG conceptualized the project. HR conducted the analysis and prepared the visualizations. OG validated and supervised the project. Both authors contributed to the writing and editing of the manuscript.  
\appendix

\section*{Supplementary Information}
% \date{}

% \maketitle

% \setcounter{page}{1}
\counterwithin{figure}{section}
\counterwithin{table}{section}

\section{Further data and methods}

\subsection{OECD data}

In addition to the validation presented using the Eora26 dataset, we replicate the analysis presented in the paper on the input-output dataset made available by the OECD.
The latest available November 2021 version of this dataset includes input-output monetary transactions between 66 countries across 45 unique industries in the time period of 1995-2018.
A list of these countries and industries is available in \autoref{supp:list_of_countries}.
The dataset is similar to the Eora26 dataset, with the exception that it has more industries but lacks the supplementary information about energy consumption and greenhouse gas emissions.

We follow the workflow as described in \autoref{sec:workflow}.
However, given that the number of countries available is much smaller in this dataset, the number of optimal clustering was found to be 3 rather than 4 for the Eora26 dataset. 
In the following section we expand on the details of the clustering pipeline used for every target industry.

\subsection{Clustering}

Here we discuss in detail the clustering pipeline, as described in \autoref{sec:clustering}, to identify countries with similar production technologies.
First, we build a feature vector for every country that includes the median marginal products of all the input industries to the target industry.
This would be a 26-dimensional feature vector for the Eora26 dataset and 45-dimensional feature vector for the OECD dataset.

Apart from the main features, we include auxiliary ones from development indicators with the aim of avoiding nonsensical clusters.
These are the median GDP per capita, median labor efficiency score and the median quality of infrastructure scores.
These medians are taken over the period of dataset used.
However, the labor efficiency score and the infrastructure score are only available for the 2007-2017 period.
Since these features are all exponentially distributed, they are log-transformed before proceeding to the clustering.

The additional features obtained from development indicators can have some multicollinearity.
Therefore, we take the first k principal components of the feature matrix which have eigenvalues greater than 1.
This reduces the dimensionality of the data to be clustered, facilitating an efficient grouping.
Taking principal components also allows us to check the importance of each feature in terms of how much it contributes to the principal component.
Thus, it helps us to ensure that the marginal product feature remains the leading one of the clustered dataset; as opposed to the auxiliary ones from development indicators.
Finally, we use consensus clustering to check the robustness of clustering and identify the optimum number of clusters k-means-constrained.

\subsection{Measuring economic sophistication}\label{sec:econ_complex}

The field of economic complexity has produced popular proxies of economic sophistication.
The two most prominent ones are the the Economic Fitness Index (EFI) \citep{tacchella_new_2012} and Economic Complexity Index (ECI) \citep{hidalgo_product_2007}.
A third one was recently develop to generalize these two and conciliate their differences: GENEPY \citep{sciarra_reconciling_2020}.
We have replicated these metrics at the country level (the most popular level of analysis in this literature) using different international trade datasets. 
For this study, we use the BACI dataset \citep{gaulier_baci_2010} that records trading data between 200 countries and over 5000 products.
These products are then classified into the industry labels provided by the input-output datasets discussed below.
For each industry, we estimate the sophistication index across each country, annually, between 1995 and 2020.
We briefly describe the measures below.

These metrics are based on a measure known as Revealed Comparative Advantage (RCA), which quantifies the global competitiveness of a country in a given product.
RCA is calculated by estimating the relative exports ($q$) of a country ($c$) of a product ($p$) in the global market as

\begin{equation}
    RCA_{c,p} = \frac{\frac{q_{c,p}}{\sum_{c'}q_{c',p}}}{\frac{\sum_{p'}q_{c,p'}}{\sum_{c',p'}q_{c',p'}}}.\label{eq:RCA}
\end{equation}

As defined in \autoref{eq:RCA}, RCA is the ratio between the fractional value of export of country $c$ of product $p$, and total exports of the country $c$ across all products.
A binary matrix $M$ can be defined as $RCA > 1$, i.e., $M_{c,p} = 1$ if a country ($c$) has a competitive advantage in that given product ($p$).
RCA can be calculated on the entire product space or for a particular industry by restricting the product space to the products of a particular industries.
This enables us to estimate industry-level complexity indices across different countries.
The EFI, ECI, and GENEPY use this binary matrix $M$ to estimate indices of economic sophistication in various ways.

As mentioned before, we generate the $M$ matrices for every industry by restricting the product space to the products of this industry.
Then we use the available open-source implementations of the three indices to estimate them yearly for the entire range of the dataset (1990-2020).
For details regarding the calculations of these metrics we refer the readers to the original papers of these metrics\citep{hidalgo_building_2009,tacchella_new_2012,sciarra_reconciling_2020}.

\begin{figure}[ht]
\centering
\caption{Economic sophistication metrics at Country and Industry level}\label{fig:complex_corr}
    \begin{minipage}{0.49\textwidth}
        \subcaption{Country level correlations}\label{fig:country.corr}
        \includegraphics[angle=0,width=1.\textwidth]{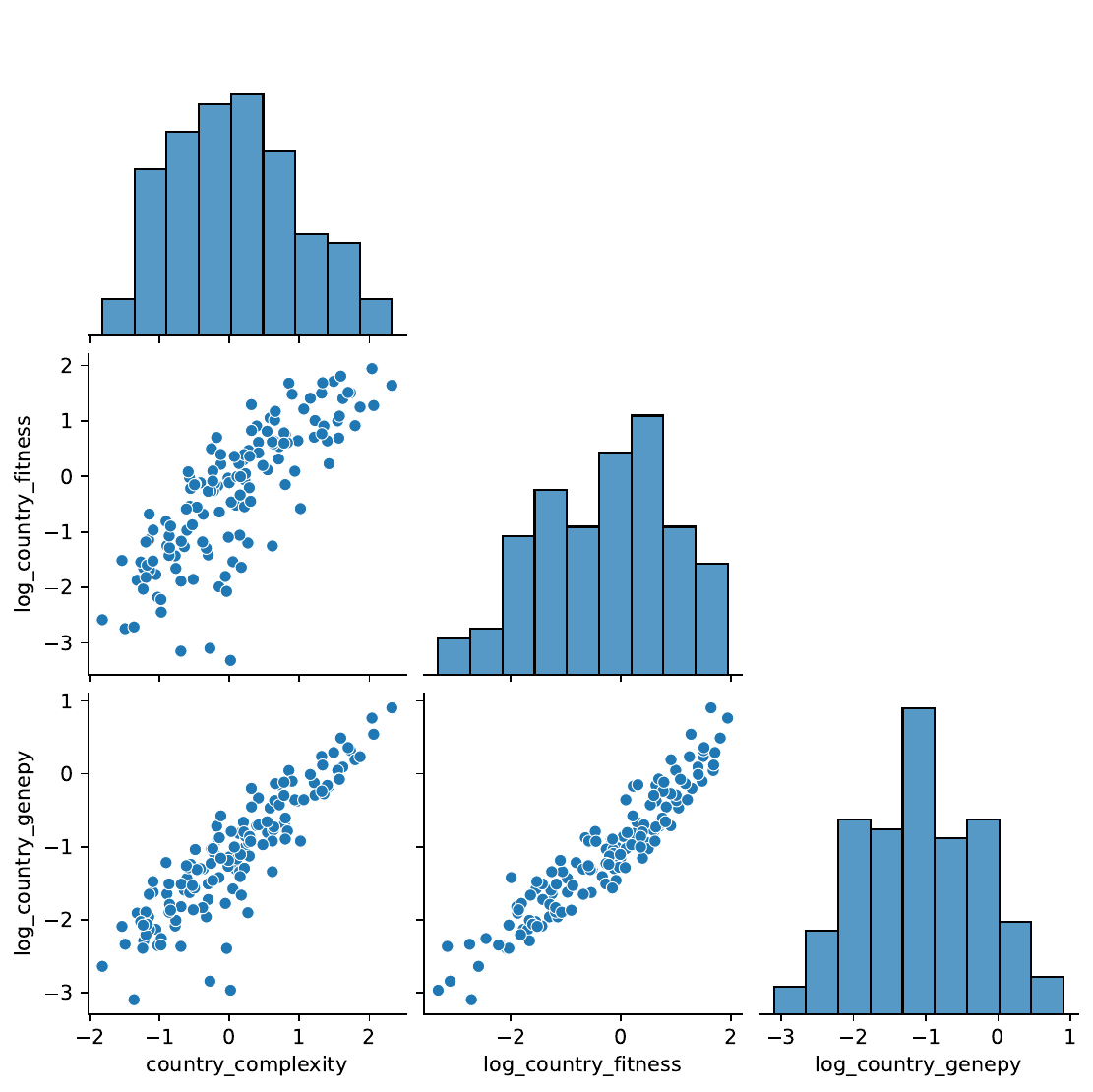}
    \end{minipage}
    \begin{minipage}{0.49\textwidth}
        \subcaption{Industry level correlations}\label{fig:industry.corr}
        \includegraphics[angle=0,width=1.\textwidth]{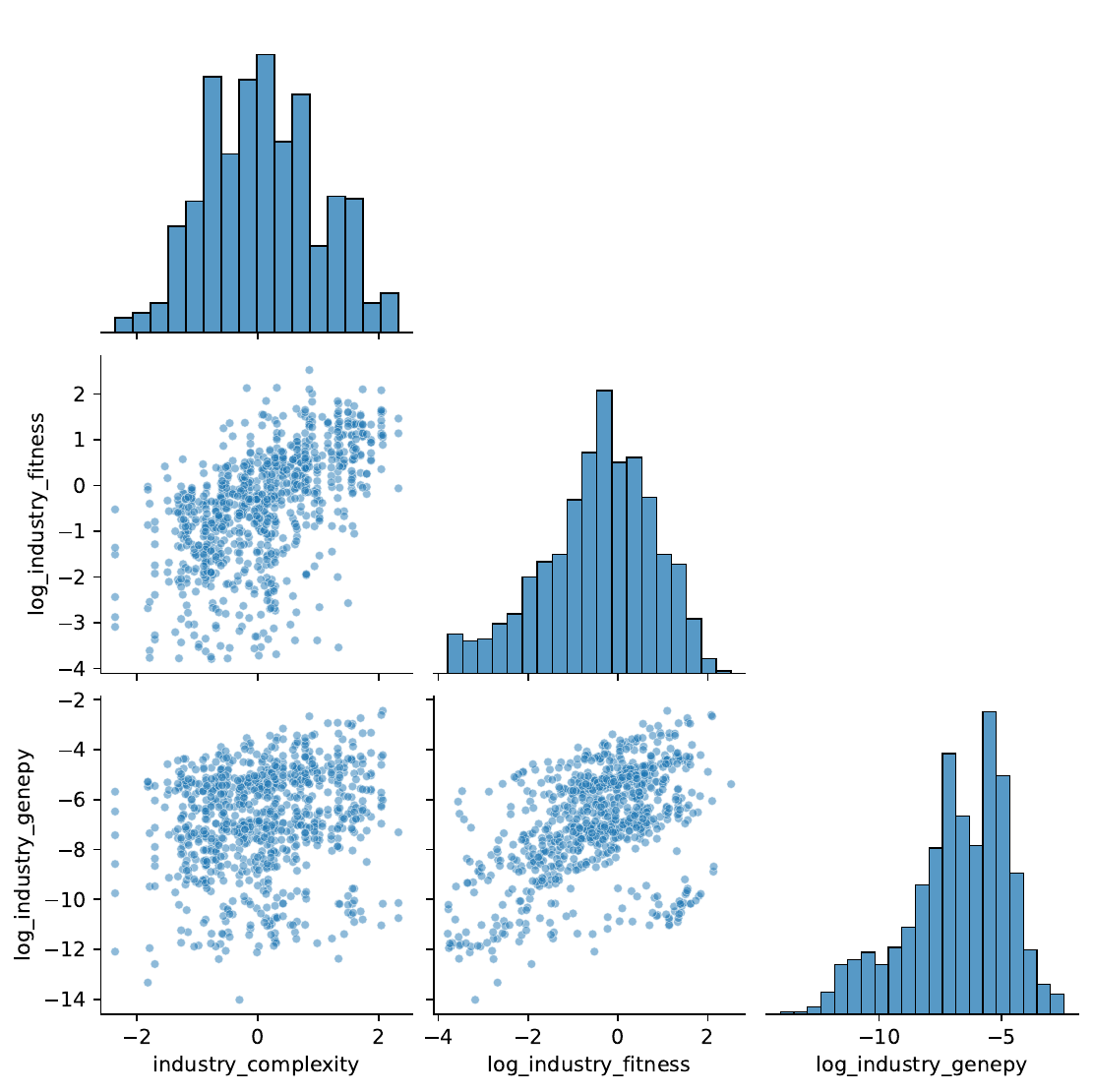}
    \end{minipage}
\end{figure}

It has been shown that these metrics are correlated when calculated at the country level \citep{sciarra_reconciling_2020}.
However, these correlations are not that strong when these measures are compared at the industry level, i.e., with the restricted product space.
The comparison between the country and the industry level indices can be seen in \autoref{fig:complex_corr}.

\hspace{5cm}

\section{Validation via regression analysis}

Let us formally test whether the synergy scores contribute, in a significant way, to the prediction of the Economic Fitness Index (EFI).
We estimate linear regression models controlling for different factors that may contribute to technological sophistication.
The intuition behind this exercise is that, if the synergy score displays a significant coefficient in the prediction of output-based complexity indices (calculated through independent methods and data), then our method is valid and quantifies the degree of technological sophistication across industries.

In \autoref{tab:reg_eora_fitness}, we present 7 linear regression models with 8 possible independent variables that contribute to the prediction of the EFI.\footnote{\autoref{fig:main_result.fitness} suggests that a linear specification is the most adequate when the EFI is the dependent variable.}
Our variable of interest is the logarithm of the synergy score.
In the first model, we estimate the association shown in \autoref{fig:main_result.fitness}; without any controls.
Model 2 includes log GDP per capita, and aims at controlling for country-specific factors such as higher income, better public governance, and better infrastructure.
Model 3 adds dummy variables indicating if the industry belongs to the primary, secondary, or tertiary sector (with the additional sector being `Other').
With this, we try to control for sector-specific factors like regulatory frameworks (e.g., tax prerogatives and unionization practices).
In model 4, we include industry-specific total output.
Finally, models 5 to 7 introduce industry-specific energy-related controls that are available in the Eora26 data.
The ratio of energy consumption to total output is a proxy of the technological efficiency of the industry, which may relate to its level of sophistication.

In all regression models, the coefficient of log synergy remains positive and statistically significant.
Furthermore, its magnitude is relatively stable across the seven models.
Interestingly, the coefficient is larger than the one of GDP per capita.\footnote{But one should be careful of not reading too much out of the controls' coefficients, as they serve the only purpose of accounting for potential confounders \citep{hunermund_nuisance_2020}.
Expecting specific signs and magnitudes in the coefficients of control variables is a common mistake known in epidemiology as the \textit{table 2 fallacy} \citep{westreich_table_2013} and in econometrics as the interpretation of \textit{endogenous controls} \citep{hunermund_causal_2021}.}
This result validates our conjecture of more synergistic information implying higher technological sophistication.
In the rest of the SI, we show that these results are robust when using GENEPY index and the Economic Complexity Index (ECI) as the dependent variable.\footnote{Note that the low adjusted $R^2$ does not invalidate our results, as this is quite common in cross-sectional regressions with a large number of observations.
If the data were aggregated, for example, by averaging the synergy scores of each industry (instead of using the input-level observations), then the explained variance would increase substantially.
This is consistent with \autoref{fig:main_result}, where the data have been binned.}

\vspace{.5 cm}
\begin{filecontents*}{Tabs/reg_eora_fitness.csv}
Predictor,Model 1,Model 2,Model 3,Model 4,Model 5,Model 6,Model 7
\multirow{2}*{Log synergy},$2.4570^{**}$,$2.3528^{**}$,$2.6970^{***}$,$2.1796^{***}$,$2.5636^{***}$,$2.6875^{***}$,$2.1795^{***}$
,$(1.0484)$,$(1.0053)$,$(1.0254)$,$(0.7582)$,$(0.9219)$,$(1.0247)$,$(0.7591)$
\multirow{2}*{Log GDP per. cap.},,$0.6523$,$0.6403$,$-0.4684$,$0.3125$,$0.6346$,$-0.4682$
,,$(0.5745)$,$(0.5685)$,$(0.5618)$,$(0.5481)$,$(0.5681)$,$(0.5642)$
\multirow{2}*{Primary sector},,,$-5.8356^{**}$,$-7.9426^{**}$,$-6.2379^{**}$,$-5.9475^{**}$,$-7.9447^{**}$
,,,$(2.7029)$,$(3.2428)$,$(2.9617)$,$(2.6803)$,$(3.1856)$
\multirow{2}*{Secondary sector},,,$-1.4582$,$-5.7442^{**}$,$-2.0697$,$-1.5831$,$-5.7460^{**}$
,,,$(1.1045)$,$(2.3370)$,$(1.2813)$,$(1.1066)$,$(2.2913)$
\multirow{2}*{Tertiary sector},,,$-6.3603^{***}$,$-8.9321^{***}$,$-7.0664^{***}$,$-6.4499^{***}$,$-8.9336^{***}$
,,,$(2.0858)$,$(2.6226)$,$(2.0998)$,$(2.1006)$,$(2.6112)$
\multirow{2}*{Log output},,,,$1.4227^{**}$,,,$1.4223^{**}$
,,,,$(0.6162)$,,,$(0.6283)$
\multirow{2}*{Log energy},,,,,$0.3827$,,
,,,,,$(0.4706)$,,
\multirow{2}*{Energy intensity},,,,,,$-1.6006$,$-0.0405$
,,,,,,$(1.4897)$,$(1.6730)$
Adjusted $R^2$,0.0237,0.0286,0.0454,0.0891,0.0521,0.0460,0.0891
No. of observations,"695,500","695,500","695,500","695,500","695,175","695,500","695,500"
\end{filecontents*}
\DTLloaddb{reg_eora_fitness}{Tabs/reg_eora_fitness.csv}
\begin{table}[!htb]
\centering
\caption{Linear models predicting the economic fitness index of industries}\label{tab:reg_eora_fitness}
\DTLdisplaydb{reg_eora_fitness}
\caption*{\footnotesize{\textit{\textbf{Notes}}: OLS regression coefficients.
The model intercept is omitted.
The dependent variable is the EFI, calculated for each industry in the dataset, and averaged across the years in the sample period.
The stars denote the level of significance.
The number in parenthesis is the clustered standard error.
Confidence levels are indicated by * for 90\%, ** for 95\%, and *** for 99\%.}}
\end{table}

To further validate our regression results, we perform the same tests using the IO tables from the OECD.
\autoref{tab:reg_oecd_fitness} shows that our results hold with these data.
In fact, the magnitude of the log synergy coefficients nearly doubled with respect to the ones reported in \autoref{tab:reg_eora_fitness}.

\vspace{.5 cm}
\begin{filecontents*}{Tabs/reg_oecd_fitness.csv}
Predictor,Model 1,Model 2,Model 3,Model 4
\multirow{2}*{Log synergy},$4.3811^{***}$,$4.2163^{***}$,$4.4180^{***}$,$4.5699^{***}$
,$(1.2116)$,$(1.1707)$,$(1.1405)$,$(1.1588)$
\multirow{2}*{Log GDP per. cap.},,$0.6275$,$0.6192$,$0.3155$
,,$(0.5588)$,$(0.5560)$,$(0.5490)$
\multirow{2}*{Primary sector},,,$1.4040$,$0.4751$
,,,$(1.8543)$,$(2.0307)$
\multirow{2}*{Secondary sector},,,$-2.5265$,$-3.4749$
,,,$(2.5887)$,$(2.5931)$
\multirow{2}*{Tertiary sector},,,$1.0486$,$0.1175$
,,,$(1.8531)$,$(1.8746)$
\multirow{2}*{Log output},,,,$0.8703^{**}$
,,,,$(0.3507)$
Adjusted $R^2$,0.0297,0.0332,0.0406,0.0598
No. of observations,"1,550,494","1,550,494","1,550,494","1,550,494"
\end{filecontents*}
\DTLloaddb{reg_oecd_fitness}{Tabs/reg_oecd_fitness.csv}
\begin{table}[!htb]
\centering
\caption{Linear models predicting the economic fitness index of industries with OECD data}\label{tab:reg_oecd_fitness}
\DTLdisplaydb{reg_oecd_fitness}
\caption*{\footnotesize{\textit{\textbf{Notes}}: OLS regression coefficients.
The model intercept is omitted.
The dependent variable is the EFI, calculated for each industry in the dataset, and averaged across the years in the sample period.
The stars denote the level of significance.
The number in parenthesis is the clustered standard error.
Confidence levels are indicated by * for 90\%, ** for 95\%, and *** for 99\%.}}
\end{table}

\clearpage

\section{Validation robustness to alternative sophistication indices}

\autoref{tab:reg_eora_genepy} and \autoref{tab:reg_eora_complexity} present the results of estimating the linear regression models from \autoref{tab:reg_eora_fitness}, but using GENEPY and the ECI as the dependent variables.
In \autoref{tab:reg_eora_genepy}, the synergy score remains positive and significant in all models.
In \autoref{tab:reg_eora_complexity}, synergy scores are significant in models 1, 4, and 7.
This is not surprising as the GDP control takes most of the explanatory power, a well known issue with the ECI due to its high correlation to GDP.

\vspace{.5 cm}
\begin{filecontents*}{Tabs/reg_eora_genepy.csv}
Predictor,Model 1,Model 2,Model 3,Model 4,Model 5,Model 6,Model 7
\multirow{2}*{Log synergy},$0.7551^{***}$,$0.7244^{***}$,$0.6829^{***}$,$0.5429^{***}$,$0.6131^{***}$,$0.6815^{***}$,$0.5432^{***}$
,$(0.2235)$,$(0.2157)$,$(0.2205)$,$(0.1597)$,$(0.1937)$,$(0.2201)$,$(0.1599)$
\multirow{2}*{Log GDP per. cap.},,$0.1926$,$0.1917$,$-0.1085$,$0.0122$,$0.1908$,$-0.1094$
,,$(0.1877)$,$(0.1724)$,$(0.1718)$,$(0.1709)$,$(0.1725)$,$(0.1721)$
\multirow{2}*{Primary sector},,,$0.5407$,$-0.0298$,$0.3238$,$0.5236$,$-0.0202$
,,,$(0.6769)$,$(0.7267)$,$(0.6736)$,$(0.6776)$,$(0.7176)$
\multirow{2}*{Secondary sector},,,$-3.0722^{***}$,$-4.2327^{***}$,$-3.4010^{***}$,$-3.0913^{***}$,$-4.2247^{***}$
,,,$(0.4009)$,$(0.5852)$,$(0.3910)$,$(0.4026)$,$(0.5779)$
\multirow{2}*{Tertiary sector},,,$-0.7863$,$-1.4827^{**}$,$-1.1740^{*}$,$-0.8001$,$-1.4762^{**}$
,,,$(0.7080)$,$(0.7410)$,$(0.7107)$,$(0.7083)$,$(0.7386)$
\multirow{2}*{Log output},,,,$0.3852^{***}$,,,$0.3872^{***}$
,,,,$(0.1421)$,,,$(0.1439)$
\multirow{2}*{Log energy},,,,,$0.2057^{*}$,,
,,,,,$(0.1092)$,,
\multirow{2}*{Energy intensity},,,,,,$-0.2449$,$0.1799$
,,,,,,$(0.3220)$,$(0.3310)$
Adjusted $R^2$,0.0379,0.0452,0.1259,0.1801,0.1583,0.1262,0.1802
No. of observations,"695,500","695,500","695,500","695,500","695,175","695,500","695,500"
\end{filecontents*}
\DTLloaddb{reg_eora_genepy}{Tabs/reg_eora_genepy.csv}
\begin{table}[!htb]
\centering
\caption{Linear models predicting the GENEPY index of industries}\label{tab:reg_eora_genepy}
\DTLdisplaydb{reg_eora_genepy}
\caption*{\footnotesize{\textit{\textbf{Notes}}: Linear regression coefficients obtained via OLS.
The estimated intercept of the model is omitted.
The dependent variable is the GENEPY index, calculated for each industry in the dataset, and averaged across the years in the sample period.
The stars denote the level of significance and the number in parenthesis the clustered standard error.
Confidence levels are indicated by * for 90\%, ** for 95\%, and *** for 99\%.}}
\end{table}

\vspace{.5 cm}
\begin{filecontents*}{Tabs/reg_eora_complexity.csv}
Predictor,Model 1,Model 2,Model 3,Model 4,Model 5,Model 6,Model 7
\multirow{2}*{Log synergy},$0.0846^{**}$,$0.0073$,$0.0077$,$-0.0218^{*}$,$-0.0112$,$0.0083$,$-0.0214^{*}$
,$(0.0391)$,$(0.0109)$,$(0.0102)$,$(0.0118)$,$(0.0117)$,$(0.0102)$,$(0.0119)$
\multirow{2}*{Log GDP per. cap.},,$0.4843^{***}$,$0.4843^{***}$,$0.4211^{***}$,$0.4368^{***}$,$0.4846^{***}$,$0.4201^{***}$
,,$(0.0078)$,$(0.0077)$,$(0.0117)$,$(0.0112)$,$(0.0077)$,$(0.0116)$
\multirow{2}*{Primary sector},,,$-0.0106$,$-0.1307^{**}$,$-0.0683$,$-0.0041$,$-0.1209^{**}$
,,,$(0.0694)$,$(0.0618)$,$(0.0596)$,$(0.0707)$,$(0.0591)$
\multirow{2}*{Secondary sector},,,$-0.0133$,$-0.2576^{**}$,$-0.1010$,$-0.0060$,$-0.2494^{**}$
,,,$(0.0960)$,$(0.1098)$,$(0.0950)$,$(0.0970)$,$(0.1087)$
\multirow{2}*{Tertiary sector},,,$-0.0138$,$-0.1604^{***}$,$-0.1162^{**}$,$-0.0086$,$-0.1538^{***}$
,,,$(0.0702)$,$(0.0553)$,$(0.0591)$,$(0.0715)$,$(0.0521)$
\multirow{2}*{Log output},,,,$0.0811^{***}$,,,$0.0832^{***}$
,,,,$(0.0117)$,,,$(0.0117)$
\multirow{2}*{Log energy},,,,,$0.0549^{***}$,,
,,,,,$(0.0063)$,,
\multirow{2}*{Energy intensity},,,,,,$0.0937^{***}$,$0.1849^{***}$
,,,,,,$(0.0207)$,$(0.0408)$
Adjusted $R^2$,0.0062,0.6083,0.6083,0.6396,0.6394,0.6088,0.6412
No. of observations,"695,500","695,500","695,500","695,500","695,175","695,500","695,500"
\end{filecontents*}
\DTLloaddb{reg_eora_complexity}{Tabs/reg_eora_complexity.csv}
\begin{table}[!htb]
\centering
\caption{Linear models predicting the economic complexity index of industries}\label{tab:reg_eora_complexity}
\DTLdisplaydb{reg_eora_complexity}
\caption*{\footnotesize{\textit{\textbf{Notes}}: Linear regression coefficients obtained via OLS.
The estimated intercept of the model is omitted.
The dependent variable is the ECI, calculated for each industry in the dataset, and averaged across the years in the sample period.
The stars denote the level of significance and the number in parenthesis the clustered standard error.
Confidence levels are indicated by * for 90\%, ** for 95\%, and *** for 99\%.}}
\end{table}

\clearpage

\section{Validation robustness to aggregated synergy scores}

In this section, we show that the results presented in \autoref{tab:reg_eora_genepy} are robust when the pairwise synergy scores are aggregated into industry-level means.
\autoref{tab:agg_eora_fitness}, \autoref{tab:agg_eora_genepy}, and \autoref{tab:agg_eora_complexity} show that our results are almost the same as in \autoref{tab:reg_eora_fitness}, \autoref{tab:reg_eora_genepy}, and \autoref{tab:reg_eora_complexity}.
Notice how, sinde the number of observations decrease due to the aggregation, the adjusted $R^2$ increases substantially.

\vspace{.5 cm}
\begin{filecontents*}{Tabs/agg_eora_fitness.csv}
Predictor,Model 1,Model 2,Model 3,Model 4,Model 5,Model 6,Model 7
\multirow{2}*{Log synergy eig.},$0.6152^{***}$,$0.3948^{***}$,$0.4271^{***}$,$0.3016^{**}$,$0.3667^{**}$,$0.4116^{***}$,$0.2978^{**}$
,$(0.1424)$,$(0.1323)$,$(0.1409)$,$(0.1408)$,$(0.1412)$,$(0.1421)$,$(0.1421)$
\multirow{2}*{Log GDP per. cap.},,$0.3530^{***}$,$0.3445^{***}$,$0.1649^{*}$,$0.2500^{***}$,$0.3321^{***}$,$0.1643$
,,$(0.0777)$,$(0.0767)$,$(0.0979)$,$(0.0898)$,$(0.0780)$,$(0.0986)$
\multirow{2}*{Primary sector},,,$0.0081$,$-0.1795$,$-0.0479$,$-0.1062$,$-0.2311$
,,,$(0.3403)$,$(0.3289)$,$(0.3336)$,$(0.3637)$,$(0.3491)$
\multirow{2}*{Secondary sector},,,$-0.7661^{*}$,$-1.1954^{**}$,$-0.8598^{*}$,$-0.8919^{*}$,$-1.2451^{***}$
,,,$(0.4433)$,$(0.4476)$,$(0.4357)$,$(0.4656)$,$(0.4632)$
\multirow{2}*{Tertiary sector},,,$-0.1818$,$-0.3934$,$-0.2919$,$-0.2671$,$-0.4296$
,,,$(0.3583)$,$(0.3474)$,$(0.3547)$,$(0.3712)$,$(0.3584)$
\multirow{2}*{Log output},,,,$0.1692^{***}$,,,$0.1638^{**}$
,,,,$(0.0620)$,,,$(0.0635)$
\multirow{2}*{Log energy},,,,,$0.0810^{*}$,,
,,,,,$(0.0425)$,,
\multirow{2}*{Energy intensity},,,,,,$-1.7574$,$-0.8842$
,,,,,,$(1.9504)$,$(1.8845)$
Adjusted $R^2$,0.2305,0.4252,0.4534,0.5117,0.4788,0.4515,0.5045
No. of observations,60,60,60,60,60,60,60
\end{filecontents*}
\DTLloaddb{agg_eora_fitness}{Tabs/agg_eora_fitness.csv}
\begin{table}[!htb]
\centering
\caption{Aggregate linear models predicting the economic fitness index of industries}\label{tab:agg_eora_fitness}
\DTLdisplaydb{agg_eora_fitness}
\caption*{\footnotesize{\textit{\textbf{Notes}}: Linear regression coefficients obtained via OLS.
The estimated intercept of the model is omitted.
The dependent variable is the EFI, calculated for each industry in the dataset, and averaged across the years in the sample period.
The stars denote the level of significance and the number in parenthesis the standard error.
Confidence levels are indicated by * for 90\%, ** for 95\%, and *** for 99\%.}}
\end{table}

\vspace{.5 cm}
\begin{filecontents*}{Tabs/agg_eora_genepy.csv}
Predictor,Model 1,Model 2,Model 3,Model 4,Model 5,Model 6,Model 7
\multirow{2}*{Log synergy eig.},$1.4518^{***}$,$1.4733^{***}$,$1.9323^{***}$,$1.5745^{***}$,$1.7475^{***}$,$1.8755^{***}$,$1.5572^{***}$
,$(0.4844)$,$(0.5252)$,$(0.5147)$,$(0.5286)$,$(0.5211)$,$(0.5194)$,$(0.5329)$
\multirow{2}*{Log GDP per. cap.},,$-0.0344$,$-0.1382$,$-0.6506^{*}$,$-0.4275$,$-0.1839$,$-0.6533^{*}$
,,$(0.3086)$,$(0.2802)$,$(0.3674)$,$(0.3316)$,$(0.2852)$,$(0.3698)$
\multirow{2}*{Primary sector},,,$-0.2675$,$-0.8026$,$-0.4387$,$-0.6863$,$-1.0357$
,,,$(1.2436)$,$(1.2344)$,$(1.2315)$,$(1.3291)$,$(1.3090)$
\multirow{2}*{Secondary sector},,,$-4.1424^{**}$,$-5.3670^{***}$,$-4.4292^{***}$,$-4.6037^{***}$,$-5.5921^{***}$
,,,$(1.6202)$,$(1.6800)$,$(1.6085)$,$(1.7013)$,$(1.7370)$
\multirow{2}*{Tertiary sector},,,$-2.2325^{*}$,$-2.8360^{**}$,$-2.5695^{*}$,$-2.5450^{*}$,$-2.9998^{**}$
,,,$(1.3095)$,$(1.3041)$,$(1.3093)$,$(1.3566)$,$(1.3441)$
\multirow{2}*{Log output},,,,$0.4826^{**}$,,,$0.4584^{*}$
,,,,$(0.2326)$,,,$(0.2380)$
\multirow{2}*{Log energy},,,,,$0.2478$,,
,,,,,$(0.1569)$,,
\multirow{2}*{Energy intensity},,,,,,$-6.4416$,$-3.9983$
,,,,,,$(7.1275)$,$(7.0668)$
Adjusted $R^2$,0.1192,0.1039,0.2784,0.3200,0.2978,0.2759,0.3112
No. of observations,60,60,60,60,60,60,60
\end{filecontents*}
\DTLloaddb{agg_eora_genepy}{Tabs/agg_eora_genepy.csv}
\begin{table}[!htb]
\centering
\caption{Aggregate linear models predicting the GENEPY index of industries}\label{tab:agg_eora_genepy}
\DTLdisplaydb{agg_eora_genepy}
\caption*{\footnotesize{\textit{\textbf{Notes}}: Linear regression coefficients obtained via OLS.
The estimated intercept of the model is omitted.
The dependent variable is the GENEPY index, calculated for each industry in the dataset, and averaged across the years in the sample period.
The stars denote the level of significance and the number in parenthesis the standard error.
Confidence levels are indicated by * for 90\%, ** for 95\%, and *** for 99\%.}}
\end{table}

\vspace{.5 cm}
\begin{filecontents*}{Tabs/agg_eora_complexity.csv}
Predictor,Model 1,Model 2,Model 3,Model 4,Model 5,Model 6,Model 7
\multirow{2}*{Log synergy eig.},$0.3673^{***}$,$0.0175$,$0.0231$,$0.0217$,$0.0168$,$0.0262$,$0.0234$
,$(0.1193)$,$(0.0277)$,$(0.0310)$,$(0.0331)$,$(0.0319)$,$(0.0313)$,$(0.0332)$
\multirow{2}*{Log GDP per. cap.},,$0.5603^{***}$,$0.5591^{***}$,$0.5572^{***}$,$0.5493^{***}$,$0.5617^{***}$,$0.5575^{***}$
,,$(0.0163)$,$(0.0169)$,$(0.0230)$,$(0.0203)$,$(0.0172)$,$(0.0231)$
\multirow{2}*{Primary sector},,,$-0.0256$,$-0.0276$,$-0.0314$,$-0.0023$,$-0.0055$
,,,$(0.0749)$,$(0.0773)$,$(0.0754)$,$(0.0802)$,$(0.0817)$
\multirow{2}*{Secondary sector},,,$-0.0264$,$-0.0310$,$-0.0361$,$-0.0008$,$-0.0096$
,,,$(0.0976)$,$(0.1052)$,$(0.0984)$,$(0.1026)$,$(0.1084)$
\multirow{2}*{Tertiary sector},,,$-0.0349$,$-0.0372$,$-0.0463$,$-0.0176$,$-0.0216$
,,,$(0.0789)$,$(0.0817)$,$(0.0801)$,$(0.0818)$,$(0.0838)$
\multirow{2}*{Log output},,,,$0.0018$,,,$0.0041$
,,,,$(0.0146)$,,,$(0.0148)$
\multirow{2}*{Log energy},,,,,$0.0084$,,
,,,,,$(0.0096)$,,
\multirow{2}*{Energy intensity},,,,,,$0.3572$,$0.3791$
,,,,,,$(0.4299)$,$(0.4408)$
Adjusted $R^2$,0.1257,0.9592,0.9571,0.9563,0.9569,0.9569,0.9561
No. of observations,60,60,60,60,60,60,60
\end{filecontents*}
\DTLloaddb{agg_eora_complexity}{Tabs/agg_eora_complexity.csv}
\begin{table}[!htb]
\centering
\caption{Aggregate linear models predicting the economic complexity index of industries}\label{tab:agg_eora_complexity}
\DTLdisplaydb{agg_eora_complexity}
\caption*{\footnotesize{\textit{\textbf{Notes}}: Linear regression coefficients obtained via OLS.
The estimated intercept of the model is omitted.
The dependent variable is the economic complexity index, calculated for each industry in the dataset, and averaged across the years in the sample period.
The stars denote the level of significance and the number in parenthesis the standard error.
Confidence levels are indicated by * for 90\%, ** for 95\%, and *** for 99\%.}}
\end{table}

\clearpage

\section{Validation robustness to alternative clustering algorithm}

In this section, we show that our results in predicting sophistication indices are robust when employing the traditional k-means algorithms to determine the country clusters, as opposed to the constrained version.
\autoref{tab:clus_eora_fitness}, \autoref{tab:clus_eora_genepy}, and \autoref{tab:clus_eora_complexity} show very similar results to those found in \autoref{tab:reg_eora_fitness}, \autoref{tab:reg_eora_genepy}, and \autoref{tab:reg_eora_complexity}.

\vspace{.5 cm}
\begin{filecontents*}{Tabs/clus_eora_fitness.csv}
Predictor,Model 1,Model 2,Model 3,Model 4,Model 5,Model 6,Model 7
\multirow{2}*{Log synergy},$2.2407^{*}$,$2.1729^{*}$,$2.6344^{**}$,$2.1779^{**}$,$2.5267^{**}$,$2.6308^{**}$,$2.1782^{**}$
,$(1.2075)$,$(1.1738)$,$(1.2276)$,$(0.9067)$,$(1.0922)$,$(1.2268)$,$(0.9057)$
\multirow{2}*{Log GDP per. cap.},,$0.6957$,$0.6850$,$-0.4179$,$0.3576$,$0.6782$,$-0.4170$
,,$(0.4938)$,$(0.4937)$,$(0.3861)$,$(0.4425)$,$(0.4942)$,$(0.3891)$
\multirow{2}*{Primary sector},,,$-6.6138^{*}$,$-8.5713^{**}$,$-6.9876^{*}$,$-6.7389^{*}$,$-8.5821^{**}$
,,,$(3.5200)$,$(3.8999)$,$(3.7588)$,$(3.4953)$,$(3.8474)$
\multirow{2}*{Secondary sector},,,$-2.0362$,$-6.1980^{**}$,$-2.6254^{*}$,$-2.1738^{*}$,$-6.2067^{**}$
,,,$(1.2911)$,$(2.5864)$,$(1.5741)$,$(1.2725)$,$(2.5450)$
\multirow{2}*{Tertiary sector},,,$-7.0789^{***}$,$-9.5159^{***}$,$-7.7589^{***}$,$-7.1805^{***}$,$-9.5233^{***}$
,,,$(2.6277)$,$(3.1543)$,$(2.7331)$,$(2.6299)$,$(3.1468)$
\multirow{2}*{Log output},,,,$1.4090^{**}$,,,$1.4068^{**}$
,,,,$(0.6540)$,,,$(0.6666)$
\multirow{2}*{Log energy},,,,,$0.3812$,,
,,,,,$(0.5436)$,,
\multirow{2}*{Energy intensity},,,,,,$-1.7715$,$-0.1959$
,,,,,,$(1.5427)$,$(1.7352)$
Adjusted $R^2$,0.0235,0.0292,0.0487,0.0916,0.0554,0.0494,0.0916
No. of observations,"697,125","697,125","697,125","697,125","696,800","697,125","697,125"
\end{filecontents*}
\DTLloaddb{clus_eora_fitness}{Tabs/clus_eora_fitness.csv}
\begin{table}[!htb]
\centering
\caption{Linear models predicting the economic fitness index of industries using k-means clustering}\label{tab:clus_eora_fitness}
\DTLdisplaydb{clus_eora_fitness}
\caption*{\footnotesize{\textit{\textbf{Notes}}: Linear regression coefficients obtained via OLS.
The estimated intercept of the model is omitted.
The dependent variable is the EFI, calculated for each industry in the dataset, and averaged across the years in the sample period.
The stars denote the level of significance and the number in parenthesis the clustered standard error.
Confidence levels are indicated by * for 90\%, ** for 95\%, and *** for 99\%.}}
\end{table}

\vspace{.5 cm}
\begin{filecontents*}{Tabs/clus_eora_genepy.csv}
Predictor,Model 1,Model 2,Model 3,Model 4,Model 5,Model 6,Model 7
\multirow{2}*{Log synergy},$0.7137^{***}$,$0.6937^{***}$,$0.6557^{**}$,$0.5316^{**}$,$0.5983^{**}$,$0.6551^{**}$,$0.5314^{**}$
,$(0.2665)$,$(0.2610)$,$(0.2843)$,$(0.2092)$,$(0.2428)$,$(0.2842)$,$(0.2089)$
\multirow{2}*{Log GDP per. cap.},,$0.2048$,$0.2027$,$-0.0972$,$0.0224$,$0.2016$,$-0.0979$
,,$(0.1871)$,$(0.1655)$,$(0.1433)$,$(0.1507)$,$(0.1657)$,$(0.1435)$
\multirow{2}*{Primary sector},,,$0.3494$,$-0.1829$,$0.1470$,$0.3290$,$-0.1751$
,,,$(0.8772)$,$(0.9012)$,$(0.8559)$,$(0.8762)$,$(0.8928)$
\multirow{2}*{Secondary sector},,,$-3.2041^{***}$,$-4.3359^{***}$,$-3.5228^{***}$,$-3.2266^{***}$,$-4.3296^{***}$
,,,$(0.4023)$,$(0.6300)$,$(0.4165)$,$(0.4007)$,$(0.6230)$
\multirow{2}*{Tertiary sector},,,$-0.9588$,$-1.6215^{*}$,$-1.3343$,$-0.9754$,$-1.6161^{*}$
,,,$(0.8626)$,$(0.8872)$,$(0.8764)$,$(0.8591)$,$(0.8851)$
\multirow{2}*{Log output},,,,$0.3832^{**}$,,,$0.3848^{**}$
,,,,$(0.1551)$,,,$(0.1573)$
\multirow{2}*{Log energy},,,,,$0.2060$,,
,,,,,$(0.1292)$,,
\multirow{2}*{Energy intensity},,,,,,$-0.2888$,$0.1421$
,,,,,,$(0.3542)$,$(0.3783)$
Adjusted $R^2$,0.0405,0.0488,0.1273,0.1811,0.1599,0.1276,0.1812
No. of observations,"697,125","697,125","697,125","697,125","696,800","697,125","697,125"
\end{filecontents*}
\DTLloaddb{clus_eora_genepy}{Tabs/clus_eora_genepy.csv}
\begin{table}[!htb]
\centering
\caption{Linear models predicting the GENEPY index of industries using k-means clustering}\label{tab:clus_eora_genepy}
\DTLdisplaydb{clus_eora_genepy}
\caption*{\footnotesize{\textit{\textbf{Notes}}: Linear regression coefficients obtained via OLS.
The estimated intercept of the model is omitted.
The dependent variable is the GENEPY index, calculated for each industry in the dataset, and averaged across the years in the sample period.
The stars denote the level of significance and the number in parenthesis the clustered standard error.
Confidence levels are indicated by * for 90\%, ** for 95\%, and *** for 99\%.}}
\end{table}

\vspace{.5 cm}
\begin{filecontents*}{Tabs/clus_eora_complexity.csv}
Predictor,Model 1,Model 2,Model 3,Model 4,Model 5,Model 6,Model 7
\multirow{2}*{Log synergy},$0.0494$,$0.0023$,$0.0026$,$-0.0239^{**}$,$-0.0129$,$0.0028$,$-0.0242^{**}$
,$(0.0435)$,$(0.0132)$,$(0.0127)$,$(0.0112)$,$(0.0110)$,$(0.0126)$,$(0.0111)$
\multirow{2}*{Log GDP per. cap.},,$0.4836^{***}$,$0.4836^{***}$,$0.4196^{***}$,$0.4358^{***}$,$0.4840^{***}$,$0.4186^{***}$
,,$(0.0081)$,$(0.0081)$,$(0.0107)$,$(0.0116)$,$(0.0081)$,$(0.0107)$
\multirow{2}*{Primary sector},,,$-0.0113$,$-0.1250^{***}$,$-0.0653$,$-0.0048$,$-0.1147^{**}$
,,,$(0.0625)$,$(0.0463)$,$(0.0492)$,$(0.0669)$,$(0.0460)$
\multirow{2}*{Secondary sector},,,$-0.0209$,$-0.2625^{***}$,$-0.1060$,$-0.0137$,$-0.2542^{***}$
,,,$(0.0855)$,$(0.0961)$,$(0.0865)$,$(0.0888)$,$(0.0962)$
\multirow{2}*{Tertiary sector},,,$-0.0118$,$-0.1532^{***}$,$-0.1112^{*}$,$-0.0065$,$-0.1462^{***}$
,,,$(0.0714)$,$(0.0515)$,$(0.0579)$,$(0.0751)$,$(0.0506)$
\multirow{2}*{Log output},,,,$0.0818^{***}$,,,$0.0839^{***}$
,,,,$(0.0118)$,,,$(0.0117)$
\multirow{2}*{Log energy},,,,,$0.0550^{***}$,,
,,,,,$(0.0073)$,,
\multirow{2}*{Energy intensity},,,,,,$0.0929^{***}$,$0.1868^{***}$
,,,,,,$(0.0180)$,$(0.0368)$
Adjusted $R^2$,0.0025,0.6071,0.6071,0.6390,0.6385,0.6075,0.6406
No. of observations,"697,125","697,125","697,125","697,125","696,800","697,125","697,125"
\end{filecontents*}
\DTLloaddb{clus_eora_complexity}{Tabs/clus_eora_complexity.csv}
\begin{table}[!htb]
\centering
\caption{Linear models predicting the economic complexity index of industries using k-means clustering}\label{tab:clus_eora_complexity}
\DTLdisplaydb{clus_eora_complexity}
\caption*{\footnotesize{\textit{\textbf{Notes}}: Linear regression coefficients obtained via OLS.
The estimated intercept of the model is omitted.
The dependent variable is the economic complexity index, calculated for each industry in the dataset, and averaged across the years in the sample period.
The stars denote the level of significance and the number in parenthesis the clustered standard error.
Confidence levels are indicated by * for 90\%, ** for 95\%, and *** for 99\%.}}
\end{table}

\section{Validation robustness to synergy scores between input triplets}

In the main text of the paper we have discussed that the synergy scores can be estimated between pairs of input industries, or between groups of higher orders.
Computing these scores for more than two industries, however, increases the data demands substantially beyond what can be provided by IO tables.
However, with our clustering procedure we can estimate synergies between input industry triplets and verify that our results remain robust.
Thus, through tables \autoref{tab:triplets_eora_fitness}, \autoref{tab:triplets_eora_genepy}, and \autoref{tab:triplets_eora_complexity}, we show that the results presented in \autoref{tab:reg_eora_fitness}, \autoref{tab:reg_eora_genepy}, and \autoref{tab:reg_eora_complexity} hold when dealing with a higher order of synergistic interactions.

\vspace{.5 cm}
\begin{filecontents*}{Tabs/triplets_eora_fitness.csv}
Predictor,Model 1,Model 2,Model 3,Model 4,Model 5,Model 6,Model 7
\multirow{2}*{Log synergy},$3.4633^{**}$,$3.3488^{**}$,$3.8830^{***}$,$3.2941^{***}$,$3.7445^{***}$,$3.8715^{***}$,$3.2943^{***}$
,$(1.3578)$,$(1.3206)$,$(1.3434)$,$(1.0457)$,$(1.2435)$,$(1.3434)$,$(1.0476)$
\multirow{2}*{Log GDP per. cap.},,$0.5757$,$0.5507$,$-0.4824$,$0.2524$,$0.5460$,$-0.4827$
,,$(0.5528)$,$(0.5432)$,$(0.5414)$,$(0.5249)$,$(0.5431)$,$(0.5437)$
\multirow{2}*{Primary sector},,,$-6.4979^{**}$,$-8.4069^{**}$,$-6.8521^{**}$,$-6.5924^{**}$,$-8.4036^{***}$
,,,$(2.8867)$,$(3.2931)$,$(3.0992)$,$(2.8627)$,$(3.2355)$
\multirow{2}*{Secondary sector},,,$-1.5441$,$-5.5637^{**}$,$-2.1019$,$-1.6517$,$-5.5609^{**}$
,,,$(1.5565)$,$(2.3957)$,$(1.6383)$,$(1.5596)$,$(2.3513)$
\multirow{2}*{Tertiary sector},,,$-7.3526^{***}$,$-9.6580^{***}$,$-7.9763^{***}$,$-7.4270^{***}$,$-9.6558^{***}$
,,,$(2.3030)$,$(2.7244)$,$(2.3388)$,$(2.3137)$,$(2.7106)$
\multirow{2}*{Log output},,,,$1.3379^{**}$,,,$1.3386^{**}$
,,,,$(0.5803)$,,,$(0.5924)$
\multirow{2}*{Log energy},,,,,$0.3505$,,
,,,,,$(0.4501)$,,
\multirow{2}*{Energy intensity},,,,,,$-1.3807$,$0.0628$
,,,,,,$(1.4073)$,$(1.6143)$
Adjusted $R^2$,0.0427,0.0466,0.0686,0.1069,0.0740,0.0690,0.1069
No. of observations,"5,564,000","5,564,000","5,564,000","5,564,000","5,561,400","5,564,000","5,564,000"
\end{filecontents*}
\DTLloaddb{triplets_eora_fitness}{Tabs/triplets_eora_fitness.csv}
\begin{table}[!htb]
\centering
\caption{Linear models predicting the economic fitness index of industries using input triplets}\label{tab:triplets_eora_fitness}
\DTLdisplaydb{triplets_eora_fitness}
\caption*{\footnotesize{\textit{\textbf{Notes}}: Linear regression coefficients obtained via OLS.
The estimated intercept of the model is omitted.
The dependent variable is the economic fitness index, calculated for each industry in the dataset, and averaged across the years in the sample period.
The stars denote the level of significance and the number in parenthesis the clustered standard error.
Confidence levels are indicated by * for 90\%, ** for 95\%, and *** for 99\%.}}
\end{table}

\vspace{.5 cm}
\begin{filecontents*}{Tabs/triplets_eora_genepy.csv}
Predictor,Model 1,Model 2,Model 3,Model 4,Model 5,Model 6,Model 7
\multirow{2}*{Log synergy},$1.0812^{***}$,$1.0478^{***}$,$1.0276^{***}$,$0.8688^{***}$,$0.9529^{***}$,$1.0261^{***}$,$0.8695^{***}$
,$(0.2711)$,$(0.2652)$,$(0.2768)$,$(0.2102)$,$(0.2505)$,$(0.2765)$,$(0.2105)$
\multirow{2}*{Log GDP per. cap.},,$0.1680$,$0.1664$,$-0.1123$,$-0.0042$,$0.1657$,$-0.1133$
,,$(0.1822)$,$(0.1674)$,$(0.1662)$,$(0.1656)$,$(0.1675)$,$(0.1665)$
\multirow{2}*{Primary sector},,,$0.3542$,$-0.1607$,$0.1546$,$0.3417$,$-0.1497$
,,,$(0.7226)$,$(0.7453)$,$(0.7099)$,$(0.7229)$,$(0.7359)$
\multirow{2}*{Secondary sector},,,$-3.0958^{***}$,$-4.1799^{***}$,$-3.4086^{***}$,$-3.1101^{***}$,$-4.1706^{***}$
,,,$(0.5355)$,$(0.6404)$,$(0.5065)$,$(0.5375)$,$(0.6337)$
\multirow{2}*{Tertiary sector},,,$-1.0638$,$-1.6856^{**}$,$-1.4223^{*}$,$-1.0737$,$-1.6784^{**}$
,,,$(0.7525)$,$(0.7653)$,$(0.7523)$,$(0.7531)$,$(0.7624)$
\multirow{2}*{Log output},,,,$0.3608^{***}$,,,$0.3631^{***}$
,,,,$(0.1349)$,,,$(0.1366)$
\multirow{2}*{Log energy},,,,,$0.1965^{*}$,,
,,,,,$(0.1041)$,,
\multirow{2}*{Energy intensity},,,,,,$-0.1834$,$0.2082$
,,,,,,$(0.2944)$,$(0.3118)$
Adjusted $R^2$,0.0705,0.0761,0.1561,0.2033,0.1855,0.1562,0.2035
No. of observations,"5,564,000","5,564,000","5,564,000","5,564,000","5,561,400","5,564,000","5,564,000"
\end{filecontents*}
\DTLloaddb{triplets_eora_genepy}{Tabs/triplets_eora_genepy.csv}
\begin{table}[!htb]
\centering
\caption{Linear models predicting the GENEPY index of industries using input triplets}\label{tab:triplets_eora_genepy}
\DTLdisplaydb{triplets_eora_genepy}
\caption*{\footnotesize{\textit{\textbf{Notes}}: Linear regression coefficients obtained via OLS.
The estimated intercept of the model is omitted.
The dependent variable is the GENEPY index, calculated for each industry in the dataset, and averaged across the years in the sample period.
The stars denote the level of significance and the number in parenthesis the clustered standard error.
Confidence levels are indicated by * for 90\%, ** for 95\%, and *** for 99\%.}}
\end{table}

\vspace{.5 cm}
\begin{filecontents*}{Tabs/triplets_eora_complexity.csv}
Predictor,Model 1,Model 2,Model 3,Model 4,Model 5,Model 6,Model 7
\multirow{2}*{Log synergy},$0.1041^{**}$,$0.0077$,$0.0085$,$-0.0275^{*}$,$-0.0129$,$0.0092$,$-0.0268$
,$(0.0502)$,$(0.0143)$,$(0.0134)$,$(0.0166)$,$(0.0160)$,$(0.0134)$,$(0.0167)$
\multirow{2}*{Log GDP per. cap.},,$0.4842^{***}$,$0.4842^{***}$,$0.4212^{***}$,$0.4370^{***}$,$0.4845^{***}$,$0.4203^{***}$
,,$(0.0078)$,$(0.0078)$,$(0.0116)$,$(0.0112)$,$(0.0078)$,$(0.0116)$
\multirow{2}*{Primary sector},,,$-0.0114$,$-0.1278^{**}$,$-0.0670$,$-0.0049$,$-0.1181^{**}$
,,,$(0.0700)$,$(0.0611)$,$(0.0596)$,$(0.0713)$,$(0.0585)$
\multirow{2}*{Secondary sector},,,$-0.0134$,$-0.2586^{**}$,$-0.1008$,$-0.0061$,$-0.2503^{**}$
,,,$(0.0964)$,$(0.1092)$,$(0.0947)$,$(0.0973)$,$(0.1082)$
\multirow{2}*{Tertiary sector},,,$-0.0151$,$-0.1557^{***}$,$-0.1141^{*}$,$-0.0100$,$-0.1493^{***}$
,,,$(0.0707)$,$(0.0545)$,$(0.0590)$,$(0.0719)$,$(0.0514)$
\multirow{2}*{Log output},,,,$0.0816^{***}$,,,$0.0836^{***}$
,,,,$(0.0115)$,,,$(0.0115)$
\multirow{2}*{Log energy},,,,,$0.0549^{***}$,,
,,,,,$(0.0062)$,,
\multirow{2}*{Energy intensity},,,,,,$0.0940^{***}$,$0.1842^{***}$
,,,,,,$(0.0207)$,$(0.0405)$
Adjusted $R^2$,0.0085,0.6083,0.6083,0.6397,0.6394,0.6088,0.6413
No. of observations,"5,564,000","5,564,000","5,564,000","5,564,000","5,561,400","5,564,000","5,564,000"
\end{filecontents*}
\DTLloaddb{triplets_eora_complexity}{Tabs/triplets_eora_complexity.csv}
\begin{table}[!htb]
\centering
\caption{Linear models predicting the economic complexity index of industries using input triplets}\label{tab:triplets_eora_complexity}
\DTLdisplaydb{triplets_eora_complexity}
\caption*{\footnotesize{\textit{\textbf{Notes}}: Linear regression coefficients obtained via OLS.
The estimated intercept of the model is omitted.
The dependent variable is the economic complexity index, calculated for each industry in the dataset, and averaged across the years in the sample period.
The stars denote the level of significance and the number in parenthesis the clustered standard error.
Confidence levels are indicated by * for 90\%, ** for 95\%, and *** for 99\%.}}
\end{table}

\clearpage

\newpage
\section{Factor analysis of the network structure}

\begin{figure}
    \centering
    \caption{The factor loadings corresponding to the 14 network features.}
    \label{fig:factor_loadings}
    \includegraphics[width=0.5\textwidth]{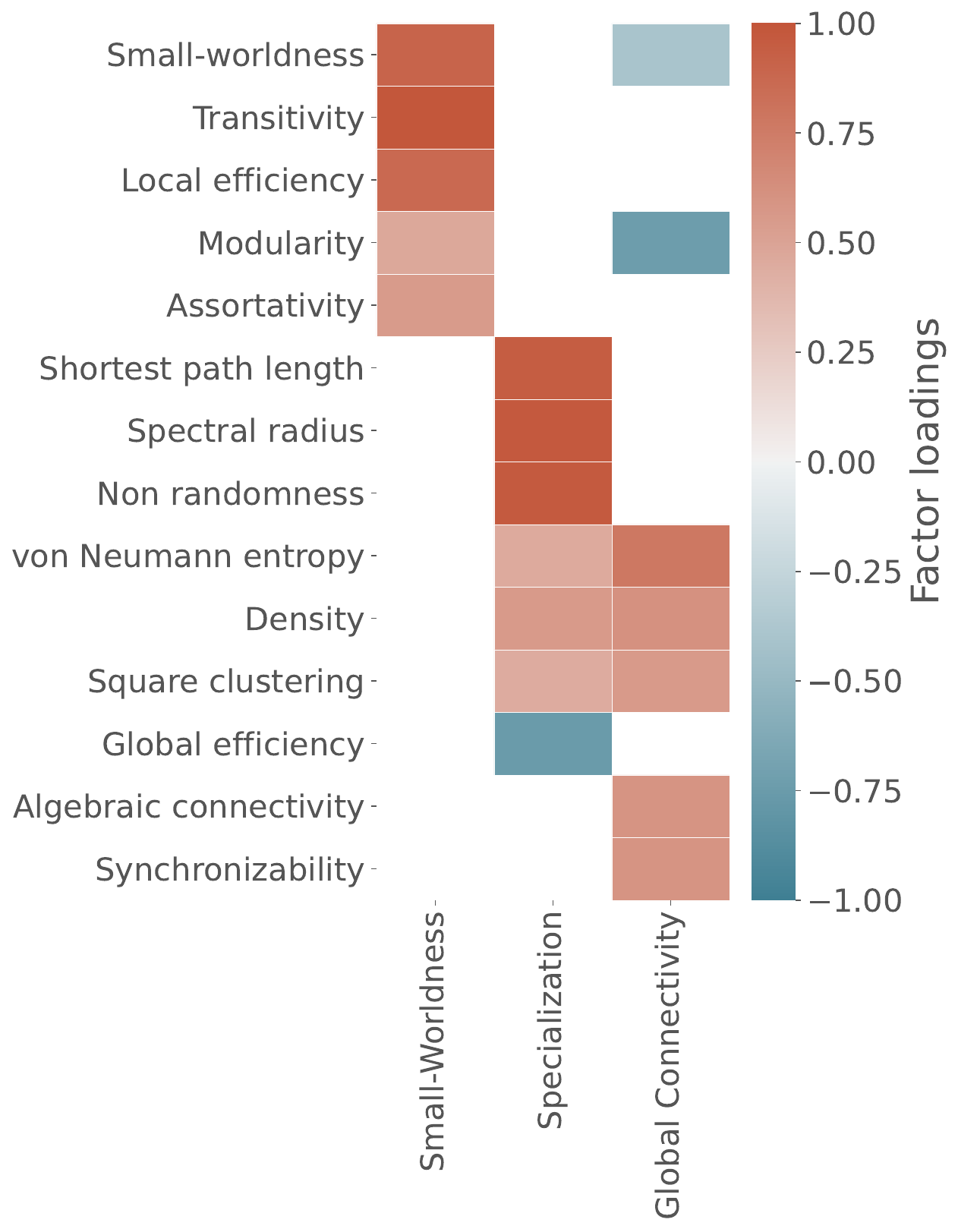}
\end{figure}

To classify the 14 global network properties used in this study, we use factor analysis to obtain three major factors characterizing synergistic production networks.
The significant feature weights of these network properties to the three factors are presented in \autoref{fig:factor_loadings}.

\newpage
\section{Validation of synergy networks using normalized flows}

To further validate the relevance of our approach, we verify if it would be possible to obtain similar topological features as those from the synergy networks directly from the IO tables.
In other words, if our results depend on the information decomposition procedure they suggest that our methodology provides non-trivial novel information about the nature and structure of technological sophistication.

\begin{figure}
    \centering
    \caption{Factor analysis of the IO network of normalized flows}\label{fig:flow_factor_out}
    \begin{minipage}{0.49\textwidth}
        \subcaption{Factor loadings}\label{fig:net_factor_si}
        \includegraphics[width=\textwidth]{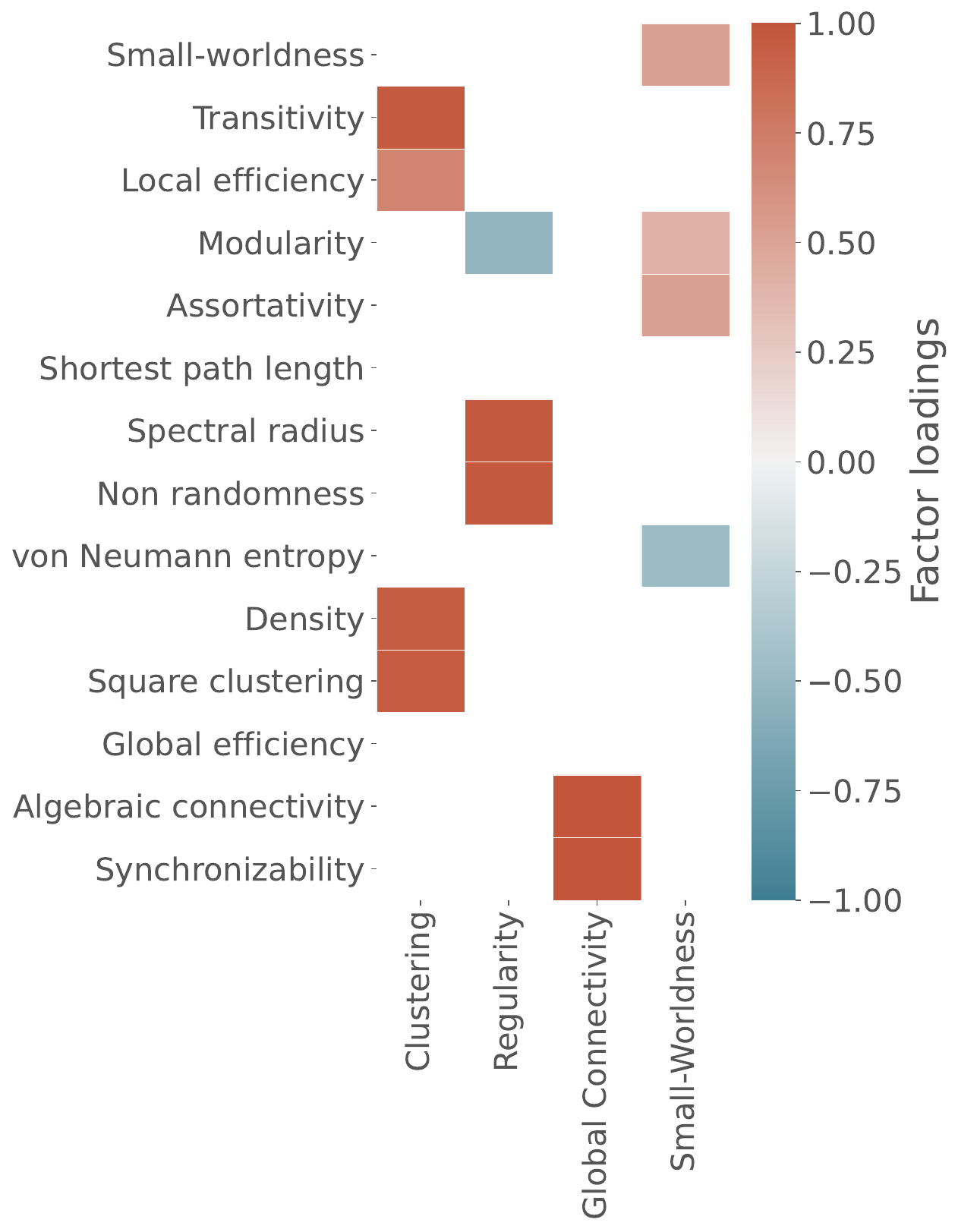}
    \end{minipage}
    \begin{minipage}{0.49\textwidth}
        \subcaption{Factor - Output correlations}\label{fig:corr_factor_flow_si}
        \includegraphics[width=\textwidth]{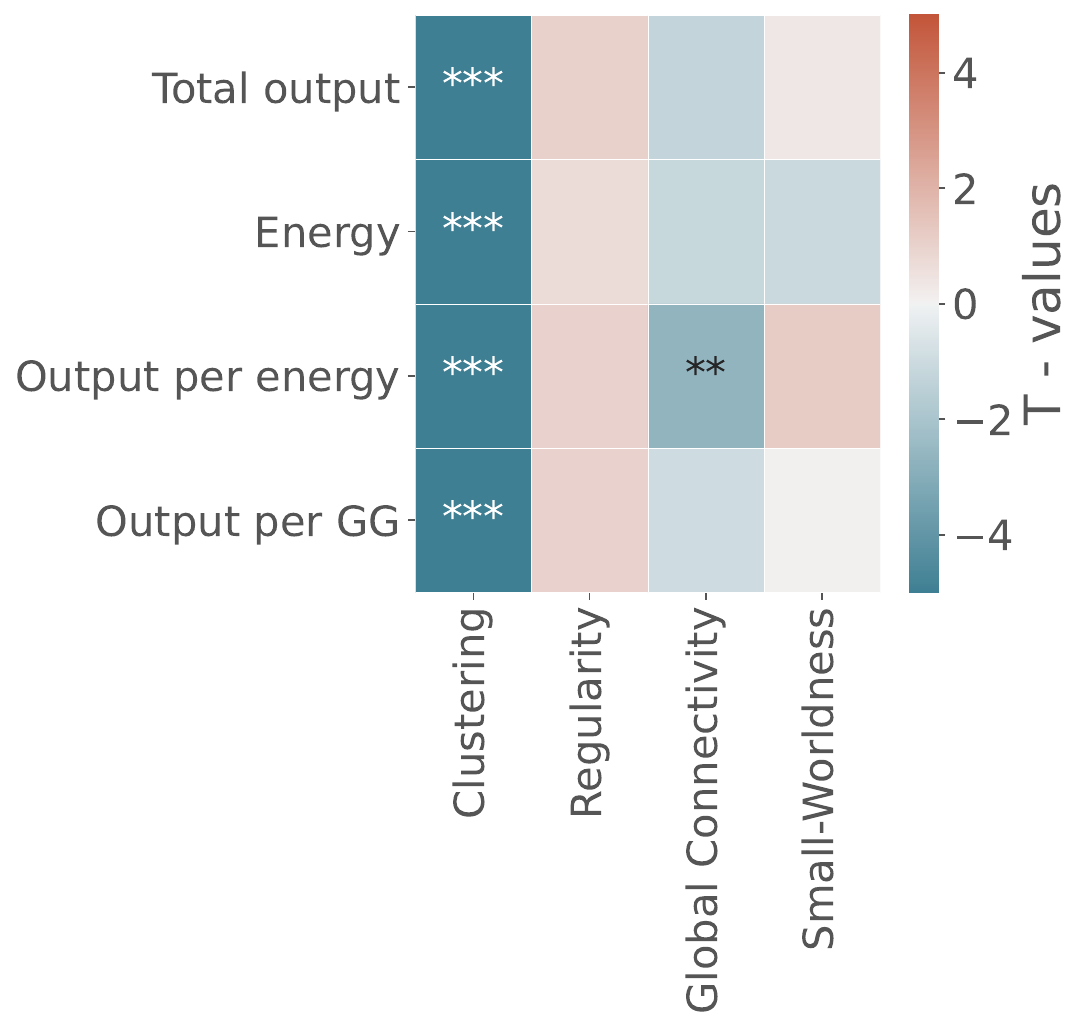}
    \end{minipage}
\end{figure}

Using the EORA dataset from the 2015 are presented here, we construct IO networks and perform the same factor analysis as in the main text (similar results are obtained for other years).
For this, we build country-level directed networks, using the flows of inputs from the source industry normalized by the total output of the target industry.
Using the same backbone algorithm on these networks, global network measures are estimated and the factor analysis is deployed.
These factors can be seen in \autoref{fig:flow_factor_out}.

In contrast with \autoref{fig:corr_factor} the factor that is closest to the small-worldness discussed in the main text has a very weak proportional contribution (about $7\%$) to the overall variance in the data, as seen in \autoref{tab:flow_fvars}.

\begin{table}[h!]
    \centering
    \caption{Factor variances associated to flow network factors}
    \label{tab:flow_fvars}
    \begin{tabular}{ccccc}
        \hline
        \textbf{Type} & \textbf{Clustering} & \textbf{Regularity} & \textbf{Global Connectivity} & \textbf{Small-Worldness}\\
        \hline
        Variance & 3.466 & 2.321 & 2.017 & 1.074 \\
        Proportional & 0.247 & 0.165 & 0.144 & 0.076 \\
        Cumulative & 0.247 & 0.413 & 0.557 & 0.634 \\
        \hline
    \end{tabular}
\end{table}

Finally, these network factors are weakly associated to output efficiency; which contrasts with our findings in the main text.
In fact the only significant relationship observed is between the clustering factor and density; which is negatively correlated with all of the metrics.
Thus, confirming the peculiar small-world structure with a connective core is unique to synergistic production networks and cannot be observed just by observing the flows between industries.

\clearpage

\section{Countries, industries, and clusters}
\label{supp:list_of_countries}

In this section, we provide all the country names, industy names, and cluster memberships in both the Eora26 and OECD datasets.

\vspace{.5 cm}
\begin{filecontents*}{Tabs/countries_eora.csv}
Code,Name
ARG,Argentina
BEL,Belgium
BOL,"Bolivia, Plurinational State of"
BRA,Brazil
CHL,Chile
CHN,China
COL,Colombia
CZE,Czechia
ECU,Ecuador
ESP,Spain
GRC,Greece
IDN,Indonesia
IND,India
IRL,Ireland
IRN,"Iran, Islamic Republic of"
ISR,Israel
KAZ,Kazakhstan
KEN,Kenya
KGZ,Kyrgyzstan
KOR,"Korea, Republic of"
KWT,Kuwait
MEX,Mexico
MLT,Malta
PER,Peru
PHL,Philippines
PRT,Portugal
PRY,Paraguay
ROU,Romania
RUS,Russian Federation
THA,Thailand
TUR,Turkey
UKR,Ukraine
URY,Uruguay
VEN,"Venezuela, Bolivarian Republic of"
VNM,Viet Nam
ZAF,South Africa
ARE,United Arab Emirates
AUT,Austria
AZE,Azerbaijan
BHR,Bahrain
BRN,Brunei Darussalam
CHE,Switzerland
CYP,Cyprus
DNK,Denmark
EST,Estonia
FIN,Finland
FRA,France
GBR,United Kingdom
GEO,Georgia
HKG,Hong Kong
HRV,Croatia
HUN,Hungary
ISL,Iceland
JPN,Japan
LTU,Lithuania
LUX,Luxembourg
LVA,Latvia
NLD,Netherlands
NOR,Norway
NZL,New Zealand
OMN,Oman
PAN,Panama
QAT,Qatar
SAU,Saudi Arabia
SGP,Singapore
SVK,Slovakia
TTO,Trinidad and Tobago
USA,United States
AUS,Australia
DEU,Germany
ITA,Italy
POL,Poland
SLV,El Salvador
SVN,Slovenia
SWE,Sweden
BDI,Burundi
BEN,Benin
BFA,Burkina Faso
BLZ,Belize
BRB,Barbados
BTN,Bhutan
CPV,Cabo Verde
ETH,Ethiopia
GMB,Gambia
HTI,Haiti
KHM,Cambodia
LAO,Lao People's Democratic Republic
LBR,Liberia
LSO,Lesotho
MDA,"Moldova, Republic of"
MKD,North Macedonia
MNE,Montenegro
MOZ,Mozambique
MUS,Mauritius
MWI,Malawi
NPL,Nepal
RWA,Rwanda
SLE,Sierra Leone
SRB,Serbia
SUR,Suriname
SWZ,Eswatini
SYC,Seychelles
TJK,Tajikistan
AGO,Angola
ALB,Albania
BGD,Bangladesh
BGR,Bulgaria
BIH,Bosnia and Herzegovina
CIV,Côte d'Ivoire
CMR,Cameroon
COD,"Congo, The Democratic Republic of the"
CRI,Costa Rica
DOM,Dominican Republic
DZA,Algeria
EGY,Egypt
GAB,Gabon
GHA,Ghana
GTM,Guatemala
GUY,Guyana
HND,Honduras
JAM,Jamaica
JOR,Jordan
LBN,Lebanon
LBY,Libya
LKA,Sri Lanka
MAR,Morocco
MLI,Mali
NIC,Nicaragua
PAK,Pakistan
SEN,Senegal
SYR,Syrian Arab Republic
TUN,Tunisia
UGA,Uganda
ZMB,Zambia
ZWE,Zimbabwe
ARM,Armenia
BWA,Botswana
GIN,Guinea
MNG,Mongolia
MRT,Mauritania
NAM,Namibia
TCD,Chad
CAN,Canada
MYS,Malaysia
NGA,Nigeria
MDG,Madagascar
MMR,Myanmar
YEM,Yemen
\end{filecontents*}
\DTLloaddb{countries_eora}{Tabs/countries_eora.csv}
\DTLdisplaylongdb[%
caption={List of countries in the Eora26 dataset},% 
label={tab:countries_eora},%
contcaption={List of countries in the Eora26 dataset (continued)},% foot={\em Continued overleaf},% lastfoot={}%
]{countries_eora}

\vspace{.5 cm}
\begin{filecontents*}{Tabs/industries_eora.csv}
Code,Name
0,Other Manufacturing
1,Others
2,Mining and Quarrying
3,Food and Beverages
4,Fishing
5,"Petroleum, Chemical and Non-Metallic Mineral Products"
6,Construction
7,Agriculture
8,Transport Equipment
9,Wood and Paper
10,"Electricity, Gas and Water"
11,"Education, Health and Other Services"
12,Metal Products
13,Textiles and Wearing Apparel
14,Electrical and Machinery
\end{filecontents*}
\DTLloaddb{industries_eora}{Tabs/industries_eora.csv}
\DTLdisplaylongdb[%
caption={List of industries in the Eora26 dataset},% 
label={tab:industries_eora},%
contcaption={List of industries in the Eora26 dataset (continued)},% foot={\em Continued overleaf},% lastfoot={}%
]{industries_eora}

\vspace{.5 cm}
\begin{filecontents*}{Tabs/clusters_eora.csv}
Country,0,1,2,3,4,5,6,7,8,9,10,11,12,13,14
ARG,3,3,2,3,3,4,3,2,1,3,3,3,3,3,4
BEL,3,2,2,2,4,4,2,2,3,2,3,2,3,2,3
BOL,3,3,1,3,4,,3,2,4,3,,4,4,3,4
BRA,3,3,2,3,4,4,3,2,4,3,3,3,3,3,4
CHL,3,,2,3,4,4,3,3,4,3,3,3,3,3,4
CHN,3,3,3,3,4,2,3,2,4,3,3,4,3,3,4
COL,3,3,2,3,4,4,3,2,4,3,3,3,3,3,4
CZE,3,3,3,2,2,4,2,3,3,2,3,2,2,2,3
ECU,3,3,2,3,4,4,3,2,4,3,3,3,3,3,4
ESP,3,,3,2,2,3,2,3,3,2,3,2,3,2,3
GRC,3,3,3,3,4,4,3,2,4,2,3,3,2,3,4
IDN,3,3,2,3,4,4,3,2,4,3,3,4,3,3,4
IND,3,3,1,3,4,2,3,2,4,3,2,3,3,1,4
IRL,3,2,3,2,4,4,2,3,3,2,2,2,2,2,3
IRN,3,,2,,,4,3,,,4,,,,,
ISR,3,3,3,3,2,3,3,4,4,2,2,2,2,3,4
KAZ,3,3,3,3,4,4,3,2,4,3,3,3,3,3,4
KEN,3,3,4,3,4,2,3,4,4,3,2,3,3,3,4
KGZ,3,4,4,4,3,2,3,4,4,4,4,4,3,4,2
KOR,3,3,,3,,4,3,2,3,3,3,3,3,3,3
KWT,3,3,2,2,2,4,3,2,4,3,3,3,3,2,4
MEX,3,3,2,3,4,4,3,2,4,3,3,3,3,3,4
MLT,3,2,4,3,3,4,3,3,4,3,2,2,2,3,3
PER,3,3,1,3,4,4,3,2,4,3,2,3,3,3,4
PHL,3,3,4,3,4,4,3,2,4,3,3,3,3,3,4
PRT,3,2,3,2,4,3,3,3,3,2,3,2,2,2,3
PRY,3,3,4,3,1,,3,2,4,3,4,3,3,3,2
ROU,3,4,3,3,3,4,3,2,4,3,3,4,3,3,4
RUS,3,4,3,3,4,4,3,2,2,3,3,3,3,3,4
THA,3,4,3,3,4,4,3,2,4,3,3,4,3,3,4
TUR,3,3,2,3,4,4,3,2,4,3,3,3,3,3,4
UKR,3,3,2,3,4,4,3,2,4,3,3,4,3,3,4
URY,3,3,3,3,4,3,3,2,4,2,2,3,2,3,4
VEN,3,3,2,3,4,1,3,2,4,3,3,3,3,3,4
VNM,3,3,2,3,4,4,3,2,,3,3,4,3,3,4
ZAF,3,3,3,3,4,4,3,4,1,3,2,,3,3,4
ARE,2,2,3,2,2,3,2,3,3,2,2,2,2,2,3
AUT,2,2,3,2,3,3,2,3,3,2,3,2,2,2,3
AZE,1,2,1,1,2,3,1,1,1,1,1,1,1,1,
BHR,2,2,3,2,,3,2,4,3,2,2,2,2,2,3
BRN,2,2,1,2,3,3,1,4,3,2,2,2,2,2,3
CHE,2,2,3,2,2,3,2,3,3,2,3,2,2,2,3
CYP,2,2,3,2,2,3,2,1,3,2,2,2,2,2,3
DNK,2,2,2,2,2,3,2,3,3,2,2,2,2,2,3
EST,2,2,3,2,2,3,2,3,4,2,2,2,2,2,3
FIN,2,2,3,2,4,3,2,3,3,3,2,,,2,3
FRA,2,2,3,2,4,3,2,3,3,2,2,2,3,2,3
GBR,2,2,2,2,2,3,2,3,3,2,3,2,2,2,3
GEO,4,2,3,3,3,4,3,2,4,3,3,1,3,4,4
HKG,2,2,,2,2,3,2,,3,3,3,4,2,3,3
HRV,2,2,3,2,2,3,1,1,3,2,1,2,2,1,3
HUN,2,2,3,2,,,2,3,3,2,2,2,2,2,3
ISL,2,2,2,4,2,3,2,3,3,2,2,2,2,2,3
JPN,2,2,3,3,2,3,2,3,3,2,2,2,2,2,3
LTU,2,2,3,2,2,4,2,3,,3,2,2,2,3,4
LUX,2,2,2,2,3,4,2,3,3,2,2,2,2,2,3
LVA,2,2,4,2,2,3,2,3,2,2,2,2,2,2,2
NLD,2,2,2,2,2,3,2,3,3,2,2,2,2,2,3
NOR,2,2,2,2,2,3,2,3,4,2,2,2,2,4,3
NZL,2,2,2,3,2,3,2,3,3,2,3,2,2,3,4
OMN,2,2,1,2,2,4,1,1,3,2,2,3,2,2,3
PAN,1,2,2,1,2,3,1,1,,1,1,1,1,1,
QAT,2,2,3,3,2,4,2,,3,2,3,3,2,2,3
SAU,2,2,,2,2,3,1,1,3,2,,2,2,2,3
SGP,2,2,3,2,2,4,2,3,3,3,3,2,2,3,3
SVK,2,2,3,2,3,4,2,3,3,3,2,2,3,3,3
TTO,1,2,1,1,2,3,1,4,1,1,1,3,1,2,1
USA,2,2,2,2,4,3,2,3,3,2,3,2,2,2,3
AUS,2,,3,2,,3,1,3,4,2,2,2,3,2,4
DEU,2,3,3,2,2,3,2,3,3,2,3,2,3,2,3
ITA,2,3,3,2,4,3,2,2,4,2,2,3,2,2,4
POL,2,3,3,2,2,3,2,3,4,2,2,2,2,2,4
SLV,1,1,3,1,1,1,1,1,1,1,1,3,1,1,1
SVN,2,3,3,2,3,3,2,3,3,2,2,2,2,3,4
SWE,2,3,3,2,4,3,2,3,3,2,2,2,2,2,3
BDI,4,4,4,4,3,2,4,,2,4,4,4,4,4,2
BEN,4,4,4,4,1,2,4,1,2,4,4,4,4,4,2
BFA,4,4,4,1,3,1,4,1,2,4,4,4,4,4,2
BLZ,4,4,4,4,3,2,4,4,2,4,4,4,4,4,2
BRB,4,4,4,4,3,2,4,4,2,4,4,4,4,4,2
BTN,4,4,4,4,3,2,4,,2,4,4,4,4,4,2
CPV,4,4,4,4,3,2,4,,2,4,4,4,4,4,2
ETH,4,,4,4,,,,4,,,4,4,4,4,2
GMB,4,4,4,4,,2,4,,2,4,,4,4,4,2
HTI,4,4,4,1,1,2,1,1,2,4,,1,4,1,1
KHM,1,1,4,4,1,1,4,4,1,1,4,1,1,1,1
LAO,4,4,4,4,1,2,4,4,2,4,1,1,4,4,2
LBR,4,4,4,4,3,2,4,4,2,4,4,4,4,4,2
LSO,4,4,4,4,3,2,4,4,2,4,4,4,4,4,2
MDA,4,,4,4,3,2,,,2,4,,4,4,4,2
MKD,2,4,4,3,3,4,2,2,4,4,2,4,3,3,2
MNE,4,4,4,4,3,2,4,4,2,4,4,4,4,4,2
MOZ,1,1,4,4,3,2,4,4,2,1,1,1,4,4,2
MUS,4,4,4,3,2,2,3,4,4,4,2,4,4,3,4
MWI,4,4,4,4,3,2,4,4,2,4,4,1,4,4,2
NPL,4,1,4,4,1,1,4,4,2,4,4,1,4,1,1
RWA,4,4,4,4,3,2,4,,2,4,4,4,4,4,2
SLE,4,4,4,4,3,2,4,,2,4,4,4,4,4,2
SRB,4,4,4,4,3,2,4,1,2,4,4,1,4,4,2
SUR,4,4,4,4,3,2,4,,2,4,4,4,4,4,2
SWZ,4,4,4,4,,2,4,,2,4,4,4,4,4,2
SYC,4,4,4,4,3,2,4,4,2,4,4,4,4,4,2
TJK,4,4,4,4,3,2,4,4,2,4,4,4,4,4,2
AGO,1,1,1,1,1,1,1,1,1,1,1,1,1,1,1
ALB,1,1,2,1,1,2,1,2,2,1,1,1,1,1,2
BGD,1,1,1,1,1,1,1,1,1,1,1,3,1,1,1
BGR,1,1,1,1,2,3,1,1,3,1,1,1,1,1,3
BIH,1,1,2,1,4,1,1,2,1,1,1,1,1,1,1
CIV,1,1,2,4,1,1,4,4,1,4,1,1,1,1,1
CMR,1,1,1,1,1,1,1,1,1,1,1,1,1,1,1
COD,4,1,1,1,1,2,4,4,1,4,4,1,4,4,2
CRI,1,1,2,1,2,1,1,4,1,1,1,1,1,1,1
DOM,1,1,,1,1,1,1,1,1,1,1,3,1,1,1
DZA,1,1,1,1,1,1,1,2,1,1,1,3,1,2,1
EGY,1,1,1,1,1,1,1,1,1,1,1,3,1,1,1
GAB,1,1,1,1,1,1,1,1,1,1,1,1,1,1,1
GHA,1,1,2,4,1,1,1,4,1,1,1,3,1,1,1
GTM,1,1,2,1,1,1,1,1,1,1,1,3,1,1,1
GUY,1,1,1,1,1,1,1,1,1,1,1,1,1,1,1
HND,1,1,1,1,1,1,1,1,1,1,1,1,1,1,1
JAM,1,1,2,4,1,2,1,1,2,1,1,1,1,1,1
JOR,1,1,1,1,2,1,1,1,1,1,4,4,1,2,1
LBN,1,1,1,1,1,1,1,1,1,1,1,1,1,1,1
LBY,1,1,1,1,1,1,1,1,1,1,1,3,1,1,1
LKA,1,1,2,1,1,1,1,1,1,1,1,3,1,1,1
MAR,1,1,2,1,1,1,1,1,1,1,1,3,1,1,1
MLI,4,1,1,1,1,1,4,4,2,4,1,1,4,4,2
NIC,1,1,,1,3,1,4,1,1,1,4,1,1,1,1
PAK,1,1,1,1,1,1,1,1,1,1,1,1,1,1,1
SEN,1,1,,4,1,1,,1,1,1,1,1,1,1,1
SYR,1,1,1,1,1,1,1,1,1,1,1,1,1,1,1
TUN,1,1,1,1,2,1,1,1,1,1,1,3,1,1,1
UGA,4,1,1,4,1,1,4,4,1,4,4,1,1,4,1
ZMB,4,1,1,1,1,1,1,,1,1,1,1,4,1,1
ZWE,1,1,1,1,1,1,4,4,1,1,1,1,4,1,1
ARM,4,4,1,1,3,1,1,2,2,1,1,1,1,4,2
BWA,4,4,2,4,3,2,4,4,2,4,4,1,4,4,2
GIN,4,4,1,4,3,2,4,,2,4,4,4,4,4,2
MNG,4,4,1,4,3,2,4,1,2,4,4,4,4,1,2
MRT,4,4,1,4,,2,4,,2,4,4,4,4,4,2
NAM,4,4,2,4,2,2,4,1,2,4,2,1,4,4,2
TCD,4,4,1,1,3,2,4,4,2,4,4,1,4,4,2
CAN,2,3,2,2,2,3,2,3,3,3,2,2,3,3,3
MYS,2,3,2,3,4,4,3,3,3,3,2,1,2,2,4
NGA,1,3,1,1,4,4,1,2,1,3,4,3,1,2,1
MDG,1,4,1,4,1,1,4,4,2,1,4,1,1,1,1
MMR,,,1,,,,,,,,,,,,
YEM,1,4,1,1,1,1,4,,1,1,4,3,1,4,1
\end{filecontents*}
\DTLloaddb{clusters_eora}{Tabs/clusters_eora.csv}
\DTLdisplaylongdb[%
caption={Country clusters by industry in the Eora26 dataset},% 
label={tab:clusters_eora},%
contcaption={Country clusters by industry in the Eora26 dataset (continued)},% foot={\em Continued overleaf},% lastfoot={}%
]{clusters_eora}

\vspace{.5 cm}
\begin{filecontents*}{Tabs/countries_oecd.csv}
Code,Name
AUT,Austria
CZE,Czechia
DEU,Germany
ESP,Spain
EST,Estonia
FIN,Finland
FRA,France
HRV,Croatia
HUN,Hungary
ITA,Italy
JPN,Japan
KAZ,Kazakhstan
KOR,"Korea, Republic of"
LTU,Lithuania
LVA,Latvia
POL,Poland
PRT,Portugal
ROU,Romania
SVK,Slovakia
SVN,Slovenia
TUR,Turkey
ARG,Argentina
AUS,Australia
BGR,Bulgaria
BRA,Brazil
CHL,Chile
CRI,Costa Rica
CYP,Cyprus
GRC,Greece
LUX,Luxembourg
MLT,Malta
NZL,New Zealand
BRN,Brunei Darussalam
COL,Colombia
HKG,Hong Kong
IDN,Indonesia
IND,India
ISR,Israel
KHM,Cambodia
LAO,Lao People's Democratic Republic
MAR,Morocco
MEX,Mexico
MMR,Myanmar
MYS,Malaysia
PER,Peru
RUS,Russian Federation
THA,Thailand
VNM,Viet Nam
CHN,China
DNK,Denmark
GBR,United Kingdom
IRL,Ireland
ISL,Iceland
NLD,Netherlands
NOR,Norway
SAU,Saudi Arabia
SGP,Singapore
SWE,Sweden
USA,United States
BEL,Belgium
CAN,Canada
CHE,Switzerland
ZAF,South Africa
TUN,Tunisia
PHL,Philippines
\end{filecontents*}
\DTLloaddb{countries_oecd}{Tabs/countries_oecd.csv}
\DTLdisplaylongdb[%
caption={List of countries in the OECD dataset},% 
label={tab:countries_oecd},%
contcaption={List of countries in the OECD dataset (continued)},% foot={\em Continued overleaf},% lastfoot={}%
]{countries_oecd}

\vspace{.5 cm}
\begin{filecontents*}{Tabs/industries_oecd.csv}
Code,Name
0,Manufacturing nec; repair and installation of machinery and equipment
1,"Pharmaceuticals, medicinal chemical and botanical products"
2,Other non-metallic mineral products
3,"Publishing, audiovisual and broadcasting activities"
4,"Arts, entertainment and recreation"
5,Fishing and aquaculture
6,Fabricated metal products
7,Electrical equipment
8,"Agriculture, hunting, forestry"
9,"Textiles, textile products, leather and footwear"
10,Paper products and printing
11,Rubber and plastics products
12,Chemical and chemical products
13,Basic metals
14,Coke and refined petroleum products
15,"Computer, electronic and optical equipment"
16,Other service activities
17,"Machinery and equipment, nec "
18,"Mining and quarrying, non-energy producing products"
19,"Electricity, gas, steam and air conditioning supply"
20,"Water supply; sewerage, waste management and remediation activities"
21,"Mining and quarrying, energy producing products"
22,Wood and products of wood and cork
23,Other transport equipment
24,"Motor vehicles, trailers and semi-trailers"
25,"Food products, beverages and tobacco"
\end{filecontents*}
\DTLloaddb{industries_oecd}{Tabs/industries_oecd.csv}
\DTLdisplaylongdb[%
caption={List of industries in the OECD dataset},% 
label={tab:industries_oecd},%
contcaption={List of industries in the OECD dataset (continued)},% foot={\em Continued overleaf},% lastfoot={}%
]{industries_oecd}

\begin{landscape}
\vspace{.5 cm}
\begin{filecontents*}{Tabs/clusters_oecd.csv}
Country,0,1,2,3,4,5,6,7,8,9,10,11,12,13,14,15,16,17,18,19,20,21,22,23,24,25
AUT,3,,2,1,1,2,2,2,2,3,2,3,1,2,1,2,1,2,2,1,2,2,1,3,3,3
CZE,3,1,1,1,2,2,3,3,2,3,2,3,1,3,1,2,2,3,3,3,3,2,1,3,3,3
DEU,3,,2,1,1,1,2,2,2,3,2,3,1,2,2,2,1,2,2,2,2,2,2,3,2,2
ESP,3,1,2,2,3,2,3,3,3,3,2,3,2,3,3,2,1,2,3,2,3,2,1,3,3,3
EST,3,4,1,1,3,1,,3,2,3,3,3,2,3,1,1,2,1,3,3,3,1,3,2,2,3
FIN,3,4,2,2,2,1,2,2,2,3,2,3,2,2,1,2,2,2,2,3,3,2,2,3,3,2
FRA,3,4,2,2,2,1,2,3,3,3,2,2,2,2,1,2,1,2,2,2,2,3,1,3,3,2
HRV,3,1,1,2,3,1,3,3,3,3,2,3,1,3,2,1,1,2,1,3,3,3,3,2,1,3
HUN,3,4,1,1,2,2,3,3,3,3,2,3,2,3,2,1,,3,2,3,3,2,1,3,3,3
ITA,3,4,2,2,1,1,3,3,3,2,2,3,1,3,1,2,1,2,1,3,3,2,1,3,2,2
JPN,3,4,1,1,3,2,,3,3,2,3,2,2,1,1,3,2,1,3,3,1,2,1,3,3,2
KAZ,3,5,3,1,3,3,1,1,3,1,1,1,3,1,3,1,3,1,3,1,1,1,3,1,1,1
KOR,3,1,2,1,1,2,3,3,1,2,3,2,1,2,3,1,2,1,3,1,3,2,1,3,3,3
LTU,3,5,1,2,2,2,3,3,2,2,2,1,3,3,1,3,2,3,3,3,3,2,1,2,2,3
LVA,3,1,1,2,3,2,3,3,2,2,3,3,1,3,2,3,2,3,1,3,3,1,,1,2,3
POL,3,1,1,3,3,2,3,3,3,3,3,3,2,3,1,3,3,3,3,3,1,2,1,3,3,3
PRT,3,1,1,2,2,1,3,3,3,1,3,,2,,3,2,1,3,3,1,2,,1,2,3,3
ROU,3,1,1,3,3,2,3,3,1,,1,2,2,3,1,3,3,3,3,1,3,2,1,3,3,1
SVK,3,1,1,2,2,2,3,3,2,3,2,3,3,3,1,3,2,3,2,3,3,2,1,3,3,3
SVN,3,,,3,2,2,3,3,2,2,2,3,3,3,2,2,2,3,2,3,3,2,1,3,3,3
TUR,3,5,1,,3,3,1,3,1,1,1,1,3,3,2,3,3,3,3,1,3,1,1,1,3,1
ARG,1,1,1,1,1,1,1,1,1,1,1,1,1,1,1,1,1,1,1,1,1,1,1,1,1,1
AUS,2,1,2,1,2,2,2,2,2,2,2,2,1,2,1,2,2,2,2,2,2,1,2,2,2,2
BGR,1,1,1,3,2,1,3,3,3,1,1,1,1,3,2,3,1,3,3,3,3,1,1,1,2,3
BRA,1,1,1,3,1,3,3,1,1,1,1,1,1,3,1,1,3,1,3,2,1,1,3,3,3,1
CHL,1,1,1,3,3,2,1,3,3,1,1,2,1,1,3,3,3,3,1,1,1,1,1,1,1,3
CRI,2,1,3,2,1,1,1,1,3,2,3,2,1,3,2,1,,3,1,3,1,1,3,1,1,3
CYP,2,1,3,2,2,1,1,1,1,2,3,2,1,2,2,1,1,2,3,3,1,1,2,2,2,3
GRC,1,1,3,2,1,3,3,1,1,3,3,3,1,,3,1,1,3,3,2,1,2,3,1,1,3
LUX,2,1,2,1,1,,2,2,3,3,3,2,2,2,,2,1,3,2,2,2,,2,2,2,2
MLT,2,1,2,2,1,2,2,2,2,2,3,2,2,2,2,3,2,3,2,3,3,1,2,2,2,3
NZL,2,1,1,2,1,1,2,2,3,2,2,2,2,2,2,3,1,2,2,1,2,1,2,2,2,2
BRN,2,,3,3,2,1,2,2,3,2,2,2,3,1,3,3,,3,1,1,2,3,2,2,,2
COL,1,5,3,3,1,3,3,3,1,1,1,2,1,3,3,3,1,3,1,1,1,3,3,1,1,1
HKG,,4,3,2,2,2,1,1,2,2,3,,3,1,1,1,2,1,,3,3,,2,1,1,2
IDN,1,4,3,3,1,3,1,1,1,1,1,1,3,1,2,3,3,1,1,1,1,1,3,1,1,1
IND,1,4,3,3,3,3,1,1,1,1,1,1,3,1,3,1,3,1,1,1,1,1,3,1,1,1
ISR,1,4,3,1,1,1,1,1,3,1,3,1,3,1,3,1,1,1,1,2,2,3,3,3,1,1
KHM,1,5,3,3,3,3,1,1,1,2,1,1,3,1,2,3,3,1,1,,3,,3,1,1,1
LAO,1,5,3,3,3,3,1,1,1,1,1,1,3,1,2,3,3,1,1,1,1,3,3,1,1,1
MAR,1,5,3,2,3,3,3,1,1,1,1,1,3,,1,3,3,1,1,3,1,1,3,3,1,1
MEX,1,5,3,3,1,1,1,1,1,1,1,1,3,1,3,1,3,1,1,1,1,3,3,1,1,1
MMR,1,5,3,3,3,3,3,3,1,1,1,1,,3,1,3,3,3,3,3,3,1,3,1,2,1
MYS,1,5,3,1,1,3,1,1,1,2,1,1,,1,1,1,1,1,3,1,1,3,1,3,3,1
PER,1,5,3,1,1,3,1,1,1,1,1,1,2,1,3,1,3,1,1,1,1,1,3,1,1,1
RUS,1,5,3,1,3,1,1,1,3,1,1,1,,1,3,1,3,1,1,1,1,3,3,1,1,1
THA,1,5,3,3,3,3,1,1,1,1,1,1,,1,3,1,3,1,1,1,1,3,3,1,1,1
VNM,1,5,3,3,3,3,1,1,1,1,1,1,,1,3,1,3,1,1,1,1,3,,1,1,1
CHN,,4,1,3,3,3,3,1,1,1,1,1,1,1,1,1,3,3,3,1,3,2,3,3,3,1
DNK,2,4,2,2,2,2,2,2,2,3,3,3,1,2,3,2,2,2,2,2,2,3,2,2,2,2
GBR,2,4,2,1,2,2,2,2,3,2,2,3,2,2,3,2,2,2,2,2,2,3,2,2,2,2
IRL,2,4,2,1,2,2,2,2,2,3,3,3,2,2,2,2,2,2,2,2,2,2,2,2,2,2
ISL,2,4,,1,1,1,2,2,3,2,3,2,3,1,1,1,1,2,1,2,2,2,2,2,2,2
NLD,2,4,2,1,1,1,2,2,2,3,3,2,,2,3,2,1,2,2,2,2,3,2,2,3,2
NOR,2,4,2,2,2,1,2,2,2,3,2,,2,2,2,2,2,2,2,2,2,3,2,2,2,2
SAU,2,4,1,3,3,2,1,2,3,2,3,2,2,2,3,3,1,3,1,2,2,3,2,2,2,2
SGP,2,4,2,2,2,2,2,2,2,2,3,2,,2,3,2,2,2,,2,2,,2,2,2,2
SWE,2,4,2,3,2,1,2,2,2,3,2,3,3,2,2,2,1,2,2,2,2,2,2,2,2,2
USA,2,4,2,1,2,1,2,2,2,3,2,3,3,2,3,2,2,2,3,2,2,3,2,3,3,3
BEL,2,,2,2,1,1,2,2,2,3,2,3,2,2,1,2,2,2,2,2,3,,1,2,3,3
CAN,2,,2,,2,2,2,2,2,3,2,3,1,1,1,2,2,2,1,2,2,3,2,2,3,2
CHE,2,,2,1,1,3,2,2,2,3,3,2,3,2,2,2,1,2,2,2,2,2,2,2,3,2
ZAF,2,,1,,2,3,2,2,3,2,3,2,3,2,2,3,2,3,3,3,2,1,3,3,2,2
TUN,1,5,1,3,3,3,,3,3,2,2,1,,3,2,3,3,1,3,3,3,1,1,3,2,3
PHL,1,5,,3,3,3,3,1,1,1,1,1,2,3,2,3,3,1,3,1,1,3,3,1,1,1
\end{filecontents*}
\DTLloaddb{clusters_oecd}{Tabs/clusters_oecd.csv}
\DTLdisplaylongdb[%
caption={Country clusters by industry in the OECD dataset},% 
label={tab:clusters_oecd},%
contcaption={Country clusters by industry in the OECD dataset (continued)},% foot={\em Continued overleaf},% lastfoot={}%
]{clusters_oecd}
\end{landscape}

\end{document}